\documentclass[aip,amsmath,amssymb,reprint,floatfix,pof]{revtex4-1}
\usepackage{graphicx}
\usepackage{dcolumn}
\usepackage{bm}
\usepackage[T1]{fontenc}
\usepackage{mathptmx}
\usepackage{etoolbox}
\usepackage{verbatim}
\usepackage{xcolor}
\usepackage{ulem}

\usepackage{hyperref}
\hypersetup{colorlinks,
 linkcolor=blue,%
 citecolor=blue,%
 urlcolor=blue
}

\newcommand{\mbr}{\mathbf{r}}
\newcommand{\mbx}{\mathbf{x}}

\newcommand{\mbu}{\mathbf{u}}

\begin{document}

\title{Dynamic instabilities and turbulence of merged rotating Bose-Einstein condensates}

\author{Anirudh Sivakumar}%
\affiliation{Department of Physics, Bharathidasan University, Tiruchirappalli 620 024, Tamil Nadu, India}

\author{Pankaj Kumar Mishra}
\affiliation{Department of Physics, Indian Institute of Technology Guwahati, Guwahati 781039, Assam, India}

\author{Ahmad A. Hujeirat}
\affiliation{Interdisciplinary Center for Scientific Computing, The University of Heidelberg, 69120 Heidelberg, Germany}

\author{Paulsamy Muruganandam}%
\affiliation{Department of Physics, Bharathidasan University, Tiruchirappalli 620 024, Tamil Nadu, India}

\begin{abstract}
We present the simulation results of merging harmonically confined rotating Bose-Einstein condensates in two dimensions. Merging of the condensate is triggered by positioning the rotation axis at the trap minima and moving both condensates towards each other while slowly ramping their rotation frequency. We analyze the dynamics of the merged condensate by letting them evolve under a single harmonic trap. We systematically investigate the formation of solitonic and vortex structures in the final, unified condensate, considering both non-rotating and rotating initial states. In both cases, merging leads to the formation of solitons that decay into vortex pairs through snake instability, and subsequently, these pairs annihilate. Soliton formation and decay-induced phase excitations generate sound waves, more pronounced when the merging time is short. We witness no sound wave generation at sufficiently longer merging times that finally leads to the condensate reaching its ground state. With rotation, we notice off-axis merging (where the rotation axes are not aligned), leading to the distortion and weakening of soliton formation. The incompressible kinetic energy spectrum exhibits a Kolmogorov-like cascade [$E(k) \sim k^{-5/3}$] in the initial stage for merging condensates rotating above a critical frequency and a Vinen-like cascade [$E(k) \sim k^{-1}$] at a later time for all cases. Our findings hold potential significance for atomic interferometry, continuous atomic lasers, and quantum sensing applications.
\end{abstract}

\date{\today}
\maketitle

\section{Introduction}

Mergers of Bose-Einstein condensates (BECs) have been the topic of great interest among the ultracold community owing to their wide applications in building up many futuristic tools such as matter-wave interferometry, continuous atom lasers, atmotronics, etc.\cite{Bloch, Andrews1997, Jo2007}. Experimentally condensate mergers have been realized in the laboratory and further investigated extensively by Scherer and group\cite{scherer2007vortex}, who demonstrated a systematic formation of the vortices as a result of the controlled merging between the condensates. This observation played a pivotal role in establishing a one-to-one relation between the dynamical instability of merged condensate along with the appearance of topological defects formed via phase transitions, popularly termed the Kibble-Zurek mechanism\cite{kibble1976topology, zurek1985cosmological}. To this end, another successful experiment on the condensate merger was carried out by Chikkatur\cite{Chikkatur2002} using optical tweezers, who observed the evaporative loss of atoms in the final state of the condensate and further systematically demonstrated that the randomized initial phases could replenish BECs which further assist in developing the vortex structures in the merged state.

Several works highlight the formation of interference patterns and their properties resulting from the mergers\cite{Andrews1997, Xiong2013, Bloch, scherer2007vortex, Shin, Stock, Hui2024}. The study of the merging process and the subsequent formation of interference patterns and solitary waves has been extended to the realms of self-gravitating BECs where fundamental solitons not only form in these head-on collisions but also survive and retain their original shapes\cite{Choi2002, Cotner2016}. Further studies on collisions of self-gravitating BECs include Paredes' observation of destructive interference patterns in solitonic collisions\cite{Paredes2016} and Schwabe's studies on gravitational cooling and density oscillations in merging of solitonic cores\cite{Schwabe2016}. Recently, Nikolaieva {\it et al.}\cite{Nikolaieva2023} investigated the stability of colliding vortex solitons in self-gravitating BECs for different orientations of angular momentum. Similar to previous literature, they also reported the survival of solitons after head-on collisions. 

The merger between the condensates and the follow-up dynamical state of the merged condensate depends very much on the phases between the condensates before merging. For instance, Stickney and colleagues\cite{Stickney} demonstrated that the mergers of condensates with opposite phases exhibit excited modes in the merged condensate, leading to the generation of unstable modes. The follow-up works by Mebrahtu\cite{Mebrahtu} and Yi\cite{Yi} suggested the phase-sensitive excitations and the influence of initial phase differences. Xiong {\it et al.}\cite{Xiong2013} analyzed the effect of the relative initial phase difference on the merging dynamics. Carretero-Gonzalez\cite{CarreteroGonzalez2008} investigated the role of recombination times on the overall structure and dynamics of the merged condensate. Buchmann {\it et al.}\cite{buchmann2009role} numerically extended the experimental study of Scherer {\it et al.}\cite{scherer2007vortex} and also explored the effects of recombination/ramping time on vortex formation in a merged condensate for a given initial relative phase and conclusively demonstrated a possibility of the formation of vortex cores during mergers that very much depends on the choice of phase difference between the condensates. Scott {\it et al.}\cite{Scott} reported the formation of persistent dark solitary waves from interference patterns, which in turn were formed from counterpropagating BEC clouds. These works, however, exclude the role of phase excitations and phase differences in the soliton dynamics. Jo {\it et al.}\cite{Jo2007} reported the appearance of condensate heating and its influence on atomic numbers using external potential gradients. Apart from this, the dynamics of BECs confined in double-well potentials that exhibit a variety of complex phases have been systematically analyzed\cite{Adhikari2014, roy2022quantum, Spagnoli2017}. In recent years, the self-bound state, mainly the quantum droplet, has been considered as a prototype model to investigate the different sorts of instabilities arising due to the elastic and inelastic collision between the solitons\cite{Ferioli2019Collisions, Gangwar:2022}.

Subsequently, in the recent past, there have been several experimental and numerical investigations that focused on the role of dynamical instabilities manifest in the system as a result of merging, which eventually takes the system to a turbulent state. For instance, Kanai {\it et al.}\cite{Kanai2019, Kanai2020, kanai2018flows} considered co-axial mergers of rotating BECs with non-uniform initial phases, where they observed the formation of spiral dark-solitons in both 2D and 3D cases, which facilitated the transfer of angular momentum between condensates. This angular momentum transfer is proportional to the initial angular momentum density difference. Furthermore, they also confirmed the formation of vortices by the decay of solitons via snake instability. 
A closely related work on turbulence aspects is the recent observation of strong quantum turbulence (SQT) in BECs by Middleton-Spencer {\it et al.}\cite{middleton2023}. They characterized SQT using the fragmented density fluctuations and non-homogeneously distributed vortex loops. SQT was achieved by oscillating the external potential, nucleating solitons, and generating density waves. Notably, Escartin {\it et al.} previously observed turbulence in superfluid helium droplets\cite{Escartin2019}, reporting Kolmogorov scaling in the incompressible kinetic spectra due to vortex-antivortex ring nucleation and subsequent weak surface turbulence.

During the past few decades, research on turbulence in 2D condensates has been extensive, encompassing both decaying and forced turbulence\cite{Numasato2009, Numasato2009a, Numasato2010, Das2022}. Recently, rotational quantum turbulence has attracted significant interest due to its fascinating properties, contrasting with its non-rotating counterpart\cite{AmetteEstrada2022, Mueller2020}. While previous studies have identified dynamical instability in merged Bose-Einstein condensates (BECs), suggesting a potential connection to turbulence, a detailed characterization of turbulence generated through mergers is still lacking. Therefore, further investigation is necessary to understand how the recombination/merging time scales influence the generation of phase excitations and their manifestation as compressible sound waves. 

In this work, we consider spatially separated, harmonically confined condensates initialized with random phases and rotating at the same frequency. We move them towards each other in real time by updating the respective trap minima and the position of the rotational axis of each condensate along the $x$-axis. Once the condensates merge at the center, we let the merged condensate evolve and analyze its behaviour. We also double the final rotation frequency compared to the initial ones. We investigate the nature of quantum turbulence and its associated incompressible kinetic energy spectra for rotating and non-rotating mergers under different merging times. We highlight the formation of dark solitons and how their dynamics are affected by rotation.  Also, we examine the possibility of an adiabatic merger, where phase excitations and the formation of compressible sound waves are suppressed in both rotating and non-rotating cases. 

The paper is structured as follows. In Sec. \ref{sec:model}, we discuss the model used for the simulation and the governing equations for studying the condensate dynamics. We also mention the condensate parameters used in these simulations. Sections \ref{subsec:non-rotating} and \ref{subsec:rotating} elucidate the numerical results of merging non-rotating and rotating condensates, respectively. Finally, we summarize the results and present our conclusions and outlook in Sec. \ref{sec:conclusion}.


\section{Mean-field Model and simulation details}
\label{sec:model}

We consider a quasi-two-dimensional condensate with strong confinement along the $z$-axis with a time-dependent rotating frequency $\Omega_t(t)$. The dynamical model describing a system of two  condensates is given by the following Gross-Pitaevskii (GP) equation in the rotating frame of reference
\begin{align}
\mathrm{i} \frac{\partial\psi}{\partial t} = & \left[-\frac{1}{2}\nabla^2 + V(\mbr,t) + g_{2D} \lvert \psi \rvert^2 - \Omega_t (t) \hat{L}_z(t) \right] \psi,
\label{eq:gpe}
\end{align}
where $\psi \equiv \psi(\mathbf{r}, t)$ denotes the condensate wave function, with $\mathbf{r} \equiv (x, y)$, $\nabla^2$ is the two-dimensional Laplace operator defined as $\nabla^2 \equiv \partial_x^2 + \partial_y^2$, and $V(\mathbf{r})$ is the external potential, which includes the harmonic trap along with the central circular barrier. The nonlinear term $g_{2D} = 4 \pi aN / (\sqrt{2\pi}d_z)$ represents the interaction strength between the atoms, where $N$ is the total number of atoms, $a$ is the s-wave scattering length, and $d_z$ corresponds to the axial width of the trap.

In Eq.~(\ref{eq:gpe}), the unit length is measured in terms of the transverse harmonic oscillator length $l_{\mbox{ho}} = \sqrt{\hbar/(m \omega)}$, time is measured in units of $\omega^{-1}$, where, $\omega$ is the transverse trap frequency. For both rotating and non-rotating cases, we fix $N = 10^4$ and $g_{2D}=400$. For the above choices of parameters, we obtain a scattering length of $a \approx 3.8a_0$. The BECs should have a strong axial confinement along $z$ axis so as to restrict the dynamics within the $x-y$ plane by implementing trap frequencies $\omega_x \sim 2 \pi \times 33 \mathrm{Hz}$, $\omega_y \sim 2\pi \times 33 \mathrm{Hz}$, $\omega_z \sim 2\pi \times 1.5 \mathrm{kHz}$. This configuration provides a trap frequency of $\omega = \left(\omega_x \omega_y \omega_z \right)^{1/3} \sim 2 \pi \times 117.8 \mathrm{Hz}$, harmonic oscillator length $l_{\mbox{ho}} \approx 1 \mu$m and unit time $\omega^{-1} \approx 1.3$ms. The oscillator length along the $z$ direction would be $l_z \approx 0.28 \mu$m. Experimentally, for a condensate of $^{87}\mathrm{Rb}$ atoms, the desired scattering length can be accessible by tuning the magnetic fields utilizing the Feshbach resonance\cite{Marcelis2004, Bauer2009}. 

To demonstrate the merger of the separated harmonic traps, we utilize the potential model and merging profile prescribed by Buchmann {\it et al.}\cite{buchmann2009role}. The potential is given by  
\begin{align}
\label{eq:merge_potential}
V(\mbr, t) = \frac{1}{2} \left[ \gamma^2(\lvert x \rvert - s(t)l)^2 + \nu^2 y^2 \right],
\end{align}
where, $\gamma=1$ and $\nu=1$ are trap aspect ratios, $l=6.4$ is the initial position of the trap minimum. 
Apart from moving the traps, it is also important to move the positions of the rotational axes and also ramp the individual frequencies of the condensates as they approach each other which can be done by modifying the angular momentum operator $\hat{L}_z(t)$ and the rotating frequency $\Omega_t(t)$, respectively.
\begin{align}
\label{eq:ramp_frequency}
\Omega_t(t) = \Omega_0 (2 - s(t))
\end{align}
\begin{align}
\label{eq:move_axes}    
\hat{L}_z(t) = \begin{cases}
\left[y \partial_x - \left(x - s(t) l\right) \partial_y \right], & \mathrm{if }\, x \geq 0 \\
\left[y \partial_x - \left(x + s(t)l \right) \partial_y \right], & \mathrm{if }\, x < 0 \\
\end{cases}
\end{align}
From these equations, one can see that the rotation axes are positioned at the trap minima and are moved simultaneously in real time.  When the rotating axes of the two separated condensates approach each other, their rotating frequencies accelerate. Upon complete overlap of the trap and axes, the final rotation frequency reaches $\Omega_f$. To ensure a seamless merger of the condensates, we have consistently assumed a final rotation frequency twice that of the initial rotation frequency, i.e., $\Omega_f = 2\Omega_0$. 

\begin{figure}[!htp]
\centering
\includegraphics[width=\linewidth]{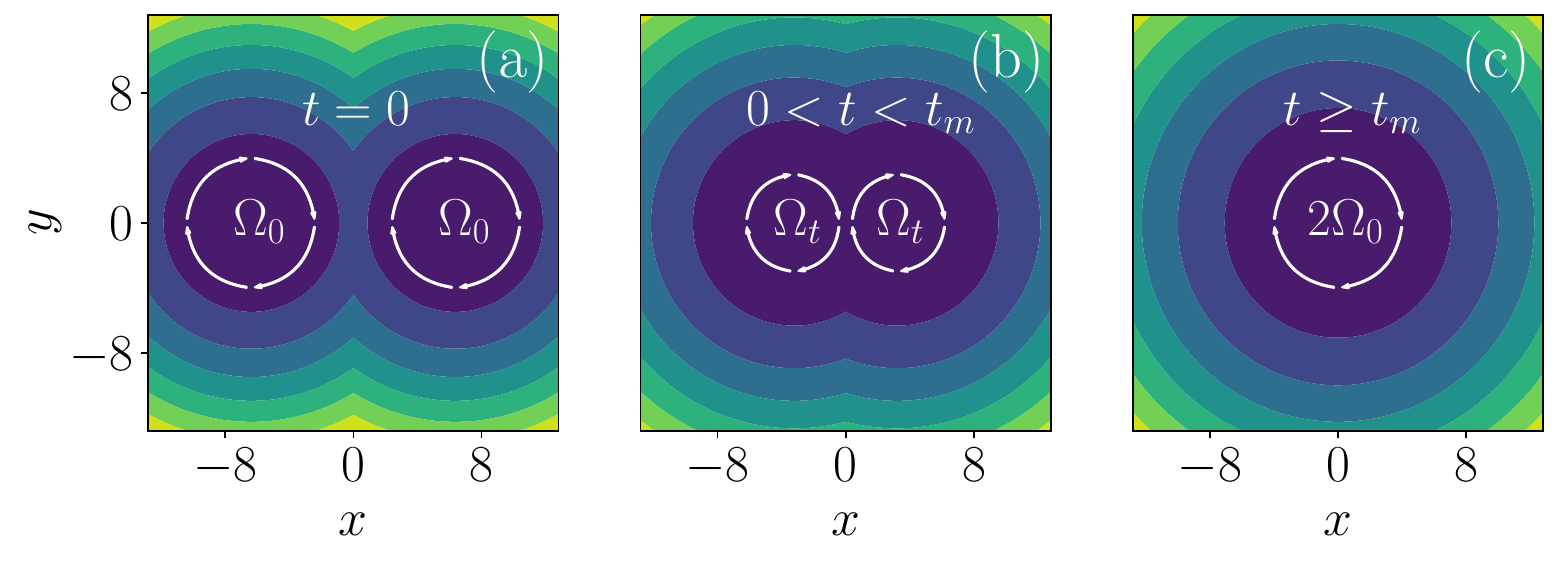}
\caption{Contour plots of the potential $V(\mathbf{r},t)$ during the merger process: (a) $t=0$, (b) $0 < t < t_m$ and (c) $t \ge t_m$. At $t=0$, the condensates are well separated and rotate at individual frequencies $\Omega_0$. As the merger progresses ($0 < t < t_m$), the potential barrier between them gradually decreases, and their respective angular frequencies become $\Omega_t$ as determined by Eq. (\ref{eq:ramp_frequency}). Once the merger is completed ($t > t_m$), the condensate is confined to a single harmonic well rotating at a frequency of $\Omega_f = 2\Omega_0$.}
\label{fig:schematic}
\end{figure}

The function $s(t)$ determines the time $t_m$ taken to perform the merger and ramp the individual rotation frequencies which are defined as
\begin{align}
s(t) = \begin{cases}
\cos^2\left(\frac{\pi t}{2 t_m}\right), & \mathrm{if\ }\, 0 \le t < t_m, \\
0, & \mathrm{if\ }\, t > t _m. 
\end{cases}
\end{align}
At $t=0$, the traps are well-separated, each rotating at a frequency $\Omega_0$ and their minima are located at $\pm l$. During the merging process, for $0 < t < t_m$, we allow the traps and the positions of their rotation axes to move towards each other. The time evolution of the trap and the axes movement is characterized by the positions of the trap minima, denoted by $\pm s(t) l$. The schematic diagrams in  Fig.~\ref{fig:schematic} display the contours and cross-sections of the potential during the merging process. %
As the condensates approach each other, the barrier between them reduces, and their respective frequencies increase and, subsequently, at the end of merging, there is a complete overlap between the two condensates now rotating with a frequency of $2\Omega_0$. In the entire duration after the complete merging, i.e., $t > t_m$, the condensate remains trapped within a single harmonic well and continues to rotate at $2 \Omega_0$. It is also noteworthy that for sufficiently large separations in the rotating case, the condensates spiral towards each other. 
We have carefully selected the parameters to minimize and avoid irrelevant dynamics in our studies. For example, we limited the initial individual rotating frequencies to $\Omega_0 \lesssim 0.5$ to ensure that the final rotating frequency ($\Omega_f$) remained below unity. Exceeding this threshold could have caused the merged condensate to escape the harmonic trap. In addition, the nonlinearity strength ($g_{2D}$) was chosen to be sufficient for vortex induction and generation of turbulence. We have also ensured identical initial rotation frequencies for both the condensate and a separation ($l$) between them that was large enough to prevent wave function overlap but not so large as to hinder spiralling effects at shorter times.

To characterize the turbulence generated in the condensate due to the complex vortex dynamics, we perform a detailed analysis of the kinetic energy spectra and corresponding fluxes at different length and time scales. For this analysis, we decompose the kinetic energy into compressible and incompressible components\cite{Nore1997} by transforming the incompressible and compressible velocity fields into $k$ space integrals via Parseval's theorem. 
\begin{figure*}[!htb]
\centering
\includegraphics[width=\linewidth]{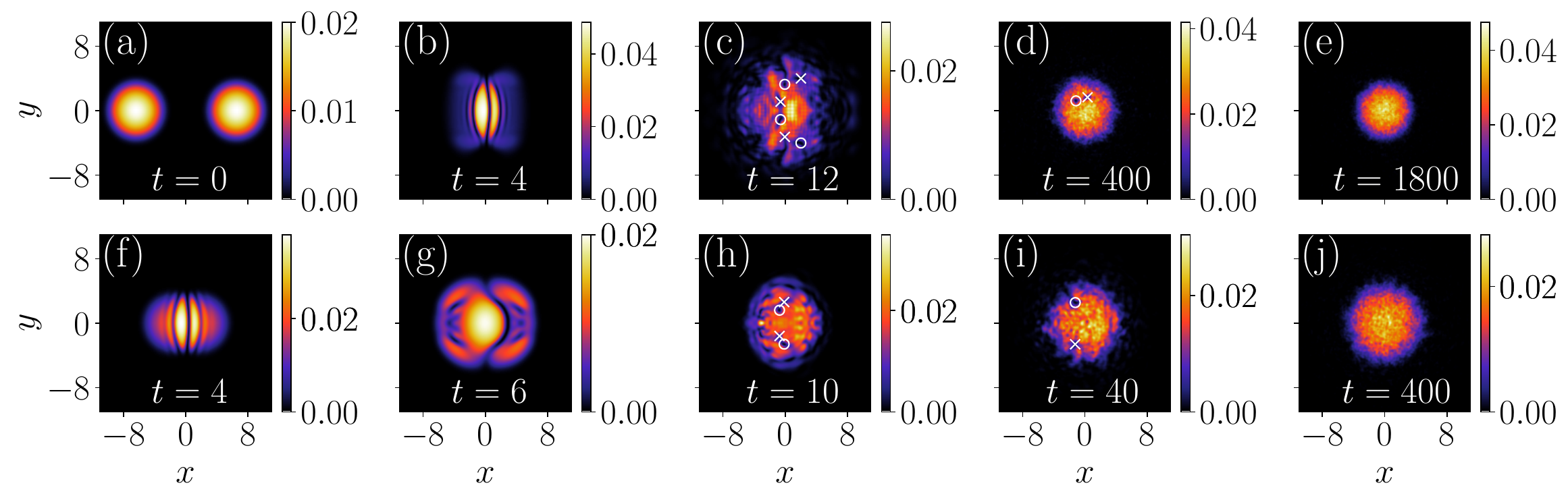}
\caption{Snapshots of the condensate density at (a) $t=0$, and at different instances for merging times $t_m=4$ [(b) - (e)] and $t_m=6$ [(f)-(j)]. (b) Formation of interference fringes alongside a central soliton, (c)-(d) shows the formation of vortex-antivortex pairs, and (e) Snapshot after the vortex pair has annihilated. (f) Formation of interference fringes, (g) soliton with significant bending, (h)-(i) formation of vortex pairs, and (j) subsequent annihilation. The anti-vortices and vortices are marked with cross ($\times$) and circle ($\circ$) symbols, respectively.}
\label{fig:nr_newdensity}
\end{figure*}%
We use the numerical implementation developed by Bradley and colleagues\cite{Bradley2022}, which evaluates these integrals using the angle-averaged Wiener-Khinchin theorem, as given below for the incompressible kinetic energy spectrum 
\begin{align}
\varepsilon_{\mathrm{kin}}^{i}(k) = \frac{m}{2}\int d^2\mbx\Lambda_2(k,\lvert\mbx\rvert)C[\mbu^{i}, \mbu^{i}](\mbx),
\label{eq:specden}
\end{align}
where $\varepsilon_{\mathrm{kin}}^{i}(k)$ is the angle-averaged incompressible kinetic energy spectrum, $\Lambda_2(k,\lvert \mbx \rvert) = (1/2\pi) k J_0(k\lvert\mbx\rvert)$ is the 2D kernel function, involving the Bessel function $J_0$ and $C[\mbu^{i}, \mbu^{i}](\mbx)$ represents the two-point auto-correlation function in position for a given incompressible velocity field. The above relation implies that for any of the position-space fields, there exists a spectral density [see Eq.~(\ref{eq:specden})] equivalent to an angle-averaged two-point correlation in $k$ space.

To quantify the detailed nature of the internal structure and associated dynamics of a merged condensate, we calculate the average force exerted by one condensate on the other. For this, we compute the condensate momentum, $\hat{P} = \psi^{*} \left( -\mathrm{i}\nabla_{\perp} \right) \psi$, where $\psi$ is the condensate wave function and $\nabla_{\perp} \equiv \left( \partial_x, \partial_y \right)$ is the two-dimensional gradient operator, which is then transformed to an angle-averaged Fourier integral. The spatial average of momentum in Fourier space is performed over the range $k=2\pi/R_{\mathrm{TF}}$ to $k=2\pi/\xi$ where $R_{\mathrm{TF}}$ and $\xi$ are the Thomas-Fermi radius and healing length of the condensate respectively.  The forcing is given as the rate of change of the spatial average of momentum $\tilde{f}(t) = \frac{d \langle \hat P \rangle}{dt}$\cite{Kanai2020}, where $\langle \cdot \rangle$ denotes the spatial average.

\section{Numerical Results}
\label{sec:numerical}

To study the merging process, we numerically solve the dynamical Gross-Pitaevskii (GP) equation \eqref{eq:gpe} using the split-Step Crank-Nicolson method\cite{muruganandam2009fortran, vudragovic2012c, young2017openmp, kumar2019c}. Our simulations are performed on a spatial grid of size $512 \times 512$ with space steps $dx = dy = 0.05$ and time step $dt = 10^{-4}$ to ensure convergence and accuracy. Similar to our previous work\cite{sivakumar2023energy}, the initial ground state consisting of two separate condensates is prepared using the imaginary time propagation scheme. Subsequently, we allow the merging of the condensates using the protocol discussed in the previous section, handled numerically using the real-time scheme for both rotating and non-rotating cases.

In the following section, we investigate different instabilities in both the rotating and non-rotating cases, including the appearance of vortices in the merged condensate, which ultimately drive the system towards turbulence. Subsequently, we characterize the turbulent fluctuations within the merged condensate by analyzing the kinetic energy spectra. Additionally, we analyze the influence of rotation on the formation and long-term evolution of vortices and solitons after the merging process and the conditions necessary for the onset of turbulence.

\subsection{Dynamical instabilities and turbulence of the merged condensate without rotation}
\label{subsec:non-rotating}
To begin with, we consider the merging dynamics of the condensate without any rotation. For this case, we prepare the initial state of two identical condensates with a random phase which helped us to take account of the effect of quantum noise present in actual experimental setups. For a systematic investigation, we consider different merging times, viz. $t_m=4$, $6$, $10$, and $20$ between the condensates and subsequently characterize their ensuing instabilities and turbulence states.
To avoid the escape of resulting condensate quickly from the single-well confinement due to uncontrolled large excitations we excluded such cases of short merging times $(t_m<4)$ . This prevents the condensate from exhibiting consistent energy spectra scaling, even for short durations. We selected $t_m$ based on the complex dynamical instabilities observed in the merged condensate, including the formation of topological defects like vortices and their subsequent chaotic dynamics leading to a turbulent state.
\begin{figure}[!htp]
\centering
\includegraphics[width=\linewidth]{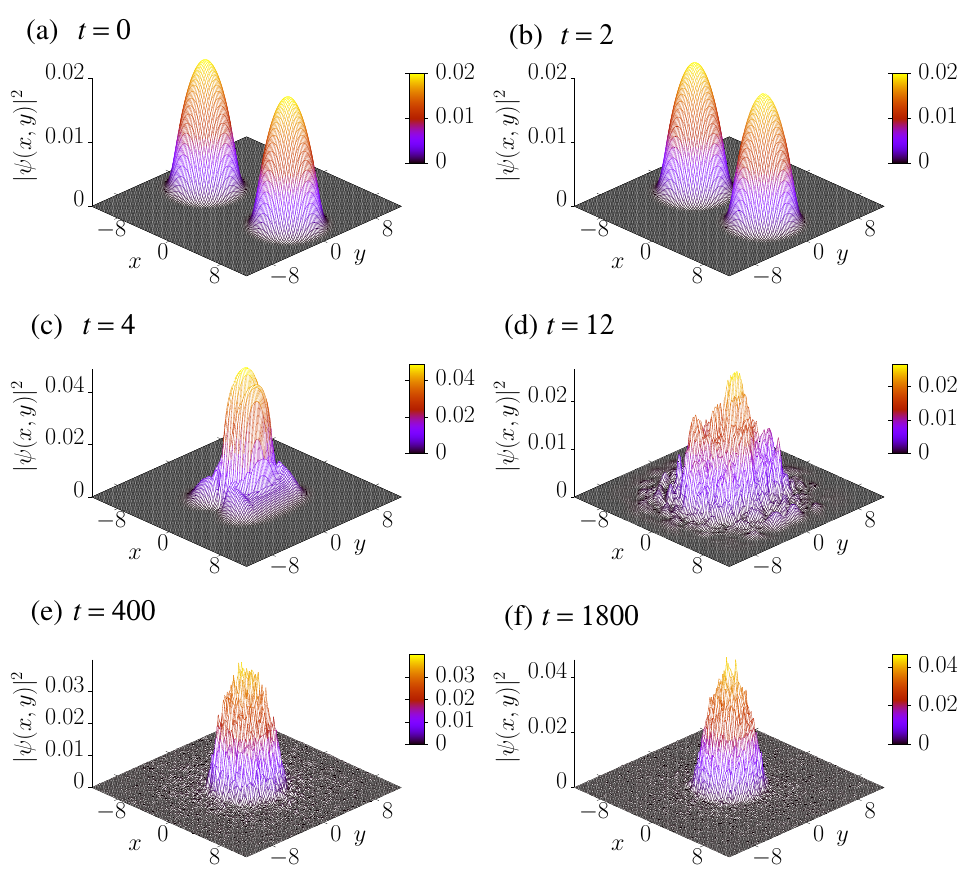}
\caption{
Three-dimensional visualizations of the condensate density surface at (a) $t=0$ [cf. Fig.~\ref{fig:nr_newdensity}(a)], (b) $t=2$, and various time instances, (c) $t=4$, (d) $t=12$, (e) $t = 400$ and (f) $t=1800$, corresponding to the merging time $t_m=4$ [cf. Fig.~\ref{fig:nr_newdensity}(b) - \ref{fig:nr_newdensity}(e)].}
\label{fig:nr:surf}
\end{figure}

Figure~\ref{fig:nr_newdensity} illustrates the evolution of condensate density profiles for merging times $t_m =4$ and $t_m=6$. The top panels (a)-(e) depict the case for $t_m=4$, while the bottom panels (f)-(j) show the case for $t_m=6$. For a more comprehensive visualization of the merging dynamics, in Fig.~\ref{fig:nr:surf}, we display the surface plots of condensate densities for $t_m=4$ corresponding to the time points depicted in Figs.~\ref{fig:nr_newdensity}(a)-\ref{fig:nr_newdensity}(e), including an intermediate time $t=2$ in Fig.~\ref{fig:nr:surf}(b). For $t_m=4$, we notice the appearance of many intricate interference fringes, particularly a centralized dark soliton [panel(b)], which remains sustained in the newly formed single harmonic trap. However, as we follow up the dynamics for a longer duration, we find that the solitonic state does not remain stable, and quickly disintegrates into the combination of the vortex-antivortex [panel (c)] pair as a result of the Snake instability\cite{verma2017generation, theocharis2003ring, mamaev1996propagation}. Later, the vortex pair annihilates, generating compressible sound waves, as shown in panel (d). Additionally, compared to the density profiles observed at a slower merging time, we also observe the presence of interference fringes for the merging time $t_m=6$ [see Figs. \ref{fig:nr_newdensity}(e)-(h)]. However, in this case, the soliton that gets generated immediately after complete merging exhibits significant bending [panel (f)], followed by a relatively quicker breakdown into vortex-antivortex pairs [panel (g)]. This particular feature triggers faster annihilation in the merged condensate compared to that of the $t_m=4$ case. For example, for $t_m=6$, the condensates show a vortexless state as a result of the vortex-antivortex annihilation even at $t\sim 400$, which was $t \sim 1800$ for $t_m=4$. The dark-soliton fringes initially formed symmetrically during the merger. However, as the merger progresses, the central soliton bends to the right and oscillates by reflecting off the trap boundaries, consistent with the equations of motion for dark solitons, as shown in Ref.\onlinecite{Frantzeskakis_2010}. 

\begin{figure}[!ht]
\centering
\includegraphics[width=\linewidth]{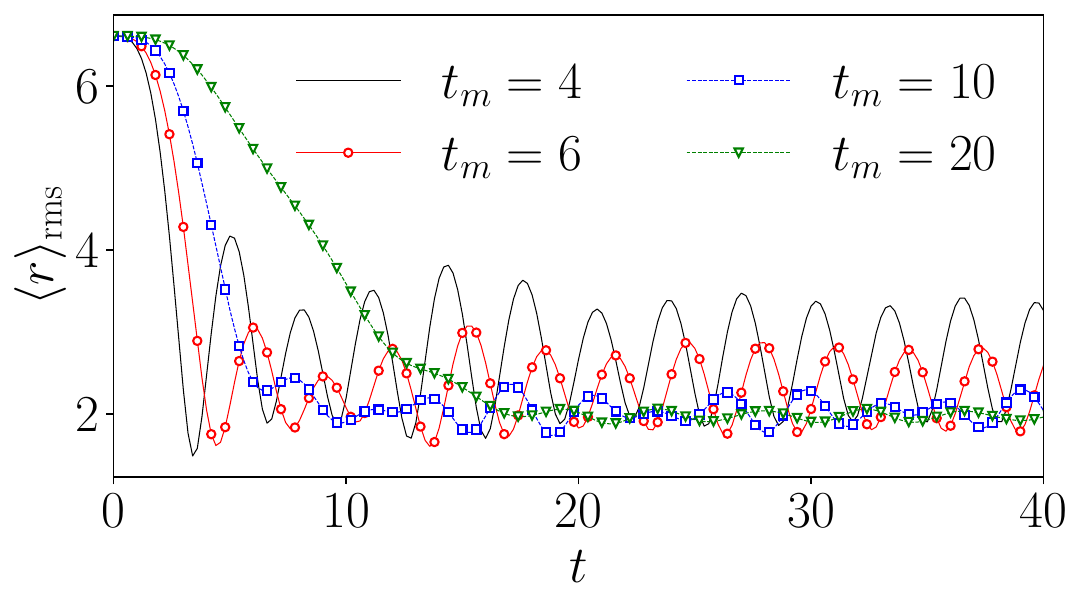}
\caption{Root mean square (rms) value of condensate radius as a function of time. The profile exhibits oscillations consistent with a frequency of $2\omega$, where $\omega$ is the trap frequency.}
\label{fig:nr_rmsplot}
\end{figure}

Our analysis reveals that the merging time significantly affects the overall dynamics and structures of the merged condensate as the condensate transitions from the double well to the single well. To effectively capture this behaviour, Fig.~\ref{fig:nr_rmsplot} presents the temporal evolution of the mean root mean squared radius of the condensate ($\langle r\rangle_{rms}$) for various merging times. %
For small merging time ($4\lesssim t_m \lesssim 6$), the merging condensate undergoes density oscillations with a frequency of $2.0\omega$. However, for longer merging time $t_m \gtrsim 6$, these oscillations subside.%
\begin{figure}[!htp]
\centering
\includegraphics[width=\linewidth]{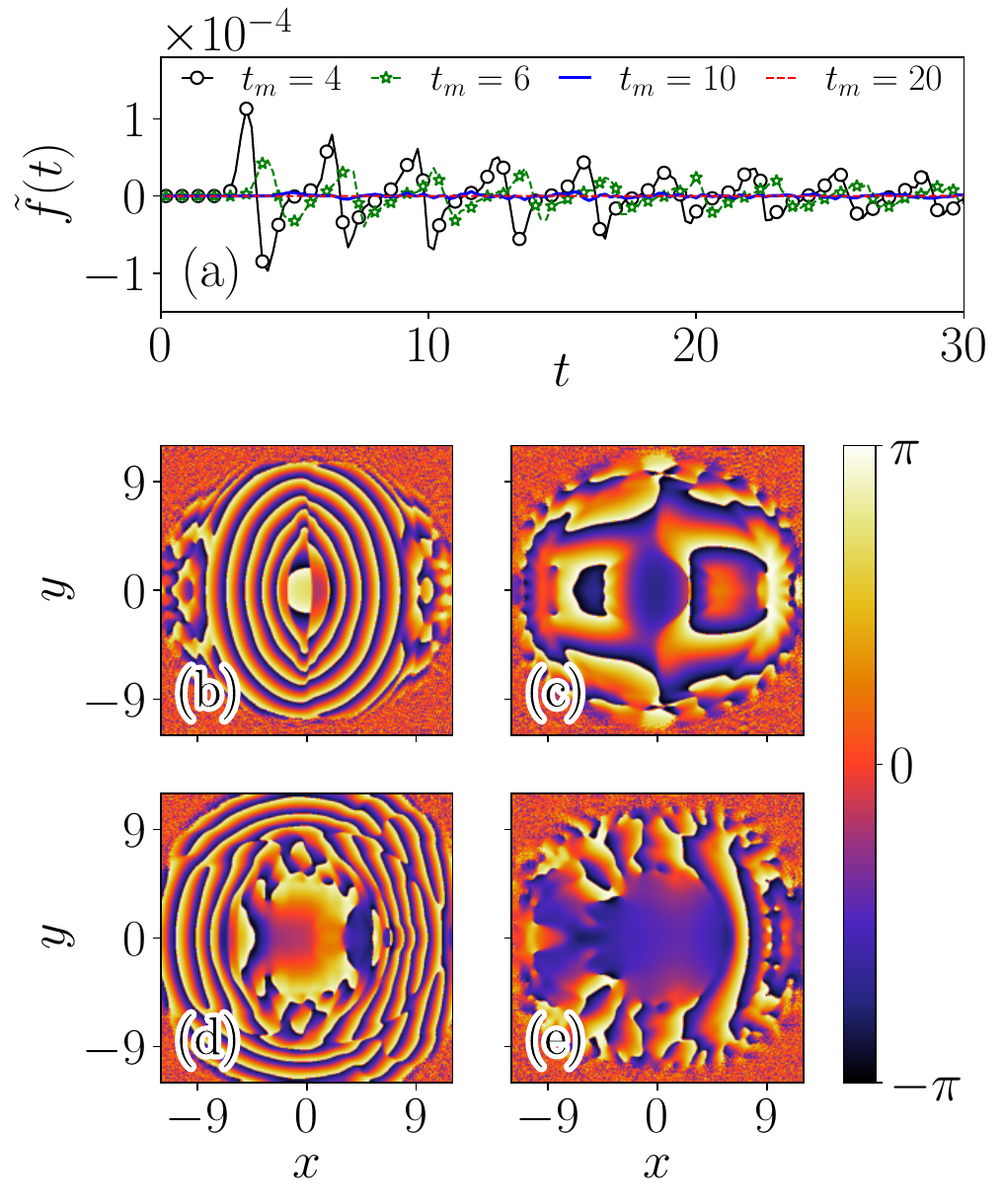}
\caption{(a) Temporal evolution of the forcing term $f$ with non-rotating case for different merging times: $t_m=4$ (black solid line with empty squares), $t_m=6$ (green line with circles), $t_m=10$ (blue solid line), and $t_m=20$ (red dashed line). Condensate phase profiles at the time of merger for various merging times (b) $t_m=4$, (c) $t_m=6$, (d) $t_m=10$ and (e) $t_m=20$.}
\label{fig:nr_forcing}
\end{figure}%

Next, we calculate the instantaneous forcing to understand the role of different competing factors involved in the instabilities of the merged condensate during various merging times. This forcing, which is the time derivative of the space-averaged density spectra, reflects the appearance of sharp discontinuities in the condensate density at a specific time. Additionally, a finite jump in force indicates the emergence of soliton formation\cite{buchmann2009role}. 
In this case, the forcing term lies within the interval $\sim [-10^{-4}, 10^{-4}]$ with the standard deviation of approximately $10^{-7}$ for $t_m=4$ and $t_m=6$, and that of the order of $10^{-8}$ for $t_m=10$ and $t_m=20$.
Figure \ref{fig:nr_forcing}(a) illustrates the time evolution of the forcing for different merging times. It shows a decrease in the forcing peak amplitude with longer merging times. This could potentially indicate phase discontinuities caused by the presence of solitons, which subsequently become less pronounced for larger merging times. The forcing peaks also decrease over time for a given merging time as the solitons eventually decay. Furthermore, the temporal evolution of the force exhibits an oscillation with a frequency of $2 \omega$ which can be attributed to the presence of the harmonic trap with frequency $\omega$. These oscillations in the forcing value can be further related to the solitons that bounce off the trap boundary and propagate in the opposite $ x$ direction. The phase profiles in Figs~\ref{fig:nr_forcing}(b)-(e) highlight the variation in soliton profiles with respect to merging time. For faster mergers, we observe a larger number of solitons with sharper phase jumps [see Fig.~\ref{fig:nr_forcing}(b)]. In contrast, slower mergers result in fewer solitons with smoother phase jumps [see Figs.~\ref{fig:nr_forcing}(c)-(e)]. The soliton velocity depends on the phase difference across the soliton\cite{Kanai2019}. A larger phase jump indicates a narrow and deep soliton moving slowly, while a smoother phase gradient signifies a wider and shallower soliton moving closer to the speed of sound.

\begin{figure}[!htp]
\centering
\includegraphics[width=\linewidth]{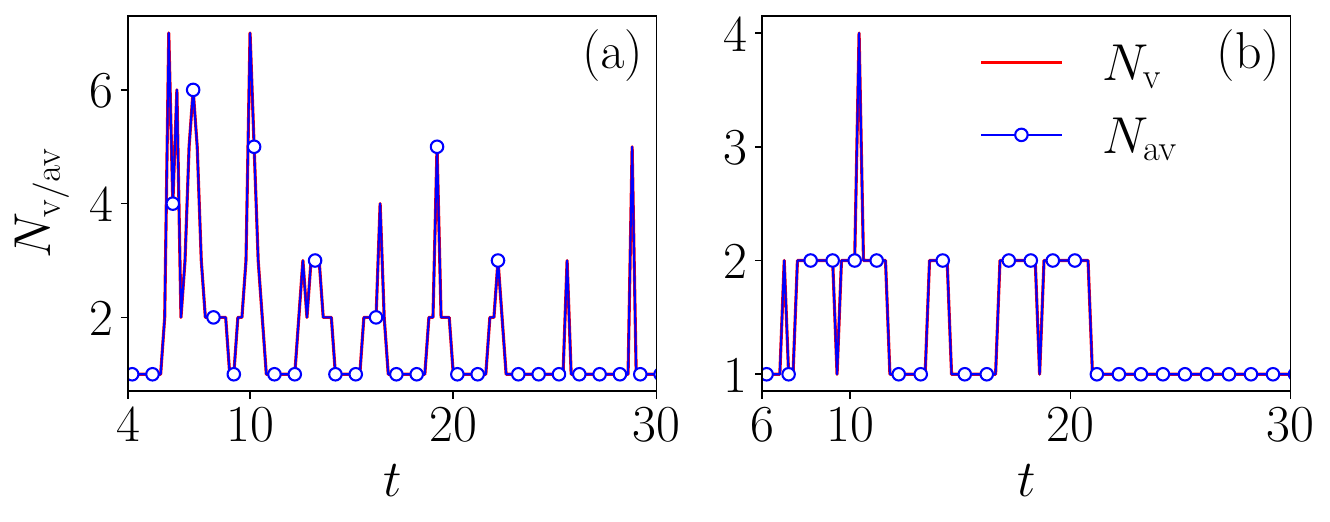}
\caption{Evolution of vortex ($N_{\mathrm{v}}$) and anti-vortex ($N_{\mathrm{av}}$) numbers within the Thomas-Fermi radius of the condensate ($R_{\mathrm{TF}}$)  for (a) $t_m=4$, (b) $t_m=6$. No vortices are observed for higher merging times.}
\label{fig:nr_vortex_number}
\end{figure}

To further analyze the mechanism responsible for the decay of solitons into vortex pairs, we investigate the temporal profile of the vortex number for $t_m=4$ and $6$, as depicted in Fig.~\ref{fig:nr_vortex_number}. We witness strong evidence of soliton decay for different merging times in the vortex numbers illustrated in Fig.~\ref{fig:nr_vortex_number}, where populations of vortices and anti-vortices appear to be equal for both merging times. The oscillations in the vortex number can be attributed to density oscillations, which are more pronounced for faster mergers  ($t_m=4$, $6$). Furthermore, evidence suggests the formation of dark neutral lumps (vortexoniums) from the coalescence of vortices and anti-vortices\cite{verma2017generation, smirnov2015scattering, groszek2016onsager}. These vortex pairs undergo a cyclical process of forming lumps and returning to vortices until the vortexonium dissipates into compressible excitations. For longer merging times, the soliton becomes wide enough to reach the sound velocity before complete condensate merging. This results in dissipation without exhibiting snake instability, which explains the absence of vortex pairs in such cases.

\begin{figure}[ht!]
\centering
\includegraphics[width=\linewidth]{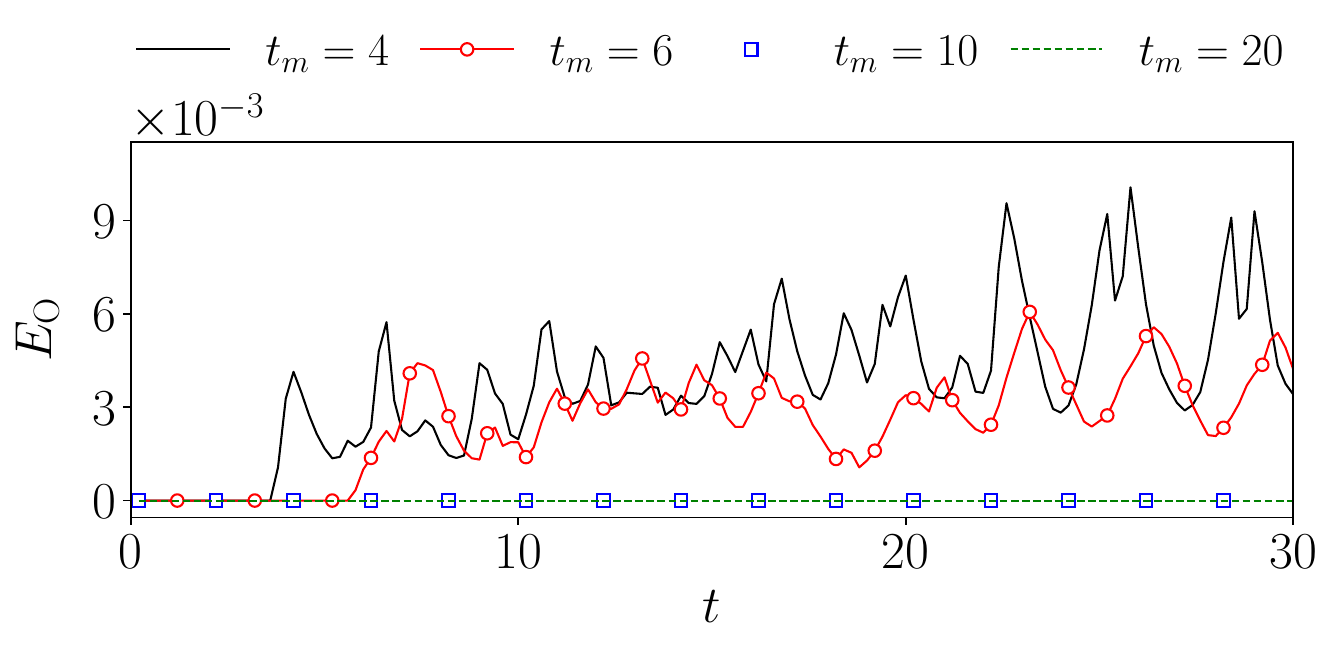}
\caption{Evolution of Onsager energy $E_{\mathrm O}$ with respect to time for the non-rotating case for merging times $t_m=4$ (black solid line), $t_m=6$ (red line with empty circles), $t_m=10$ (blue squares), $t_m=20$ (green dashed line).}
\label{fig:nr_onsager}
\end{figure}

Next, we analyze the spatial distribution of the incompressible (not shown) and compressible kinetic energies [see Fig.~\ref{fig:nr_compden} in Appendix \ref{appendix_a}] for $t_m=4$, which offer insights into the detailed nature of the fluctuations generated during the merging process and subsequent instabilities leading to the turbulent state, thereby quantifying the different characteristics of the merged condensate. The presence of point vortices contributes to the incompressible kinetic energy, whose evolution follows a similar trend as that of vortex numbers. The compressible kinetic energy also exhibits an early peak [see Fig.~\ref{fig:nr_compden}(a)-(c)] due to the trap merger generating sound waves, which eventually diminish and settle as the merged condensate evolves [see Fig.~\ref{fig:nr_compden}(d)]. Notably, the compressible kinetic energy is comparable to the incompressible kinetic energy because of the smaller number of vortices.

The Onsager energy, $E_{\mathrm O}$, that is, the energy per vortex, is also calculated and displayed in Fig.~\ref{fig:nr_onsager} for various merging times. The Onsager energy indicates that vortices formed through shorter mergers are more energetic and undergo oscillations consistent with a frequency $2.0 \omega$. 

\begin{figure}[ht!]
\centering
\includegraphics[width=\linewidth]{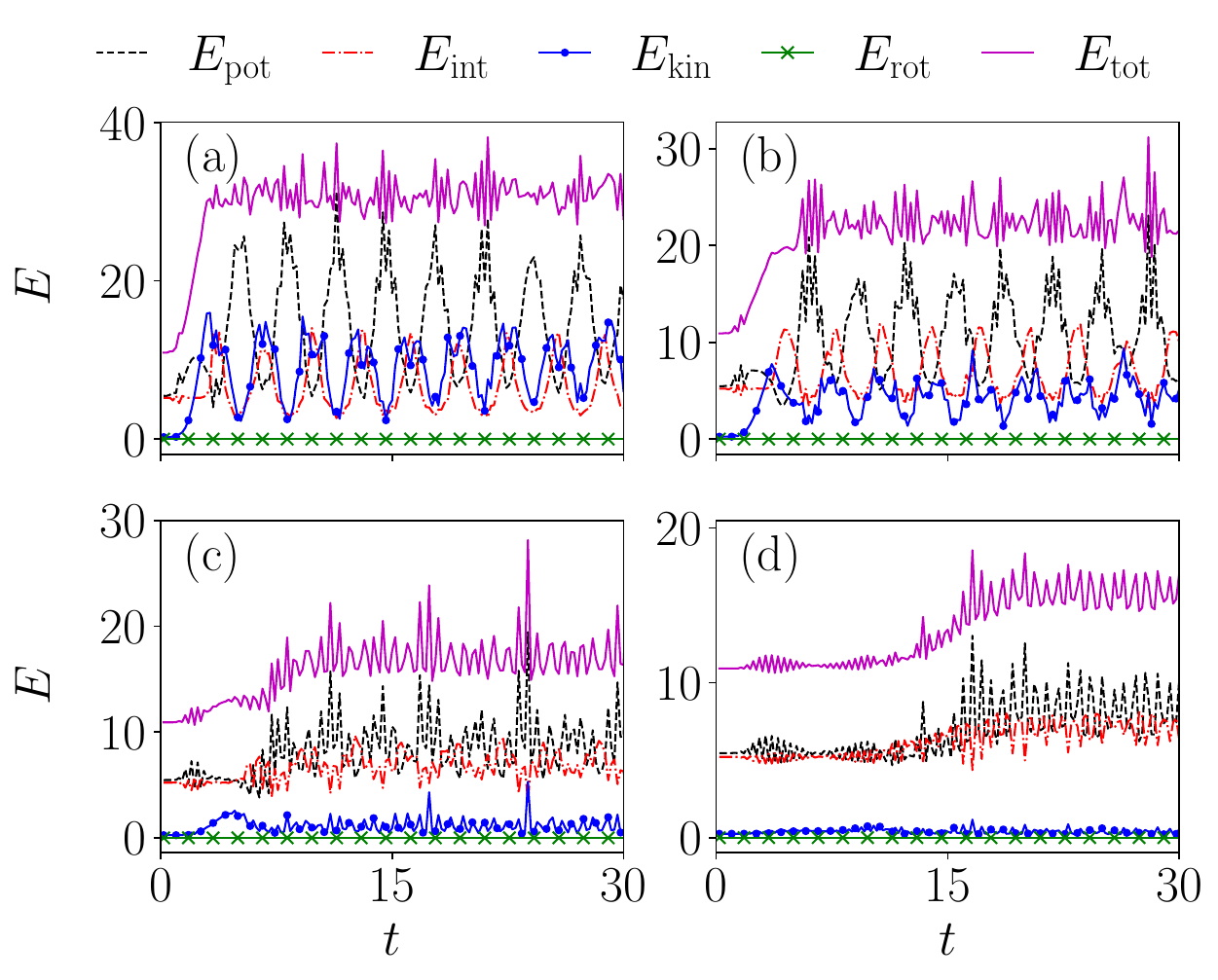}
\caption{Evolution of different components of condensate energies with respect to time for (a) $t_m=4$, (b) $t_m=6$, (c) $t_m=10$, and (d) $t_m=20$.}
\label{fig:nr_energies}
\end{figure}

Figures \ref{fig:nr_energies}(a)-(d) illustrate the temporal evolution of the various energy components, such as the expectation value of the potential energy ($E_{pot}$), interaction energy ($E_{int}$), kinetic energy ($E_{kin}$), rotational energy ($E_{rot}$) for different merging times $t_m=4$, $6$, $10$, and $20$, respectively. %
\begin{figure}[ht!]
\centering
\includegraphics[width=\linewidth]{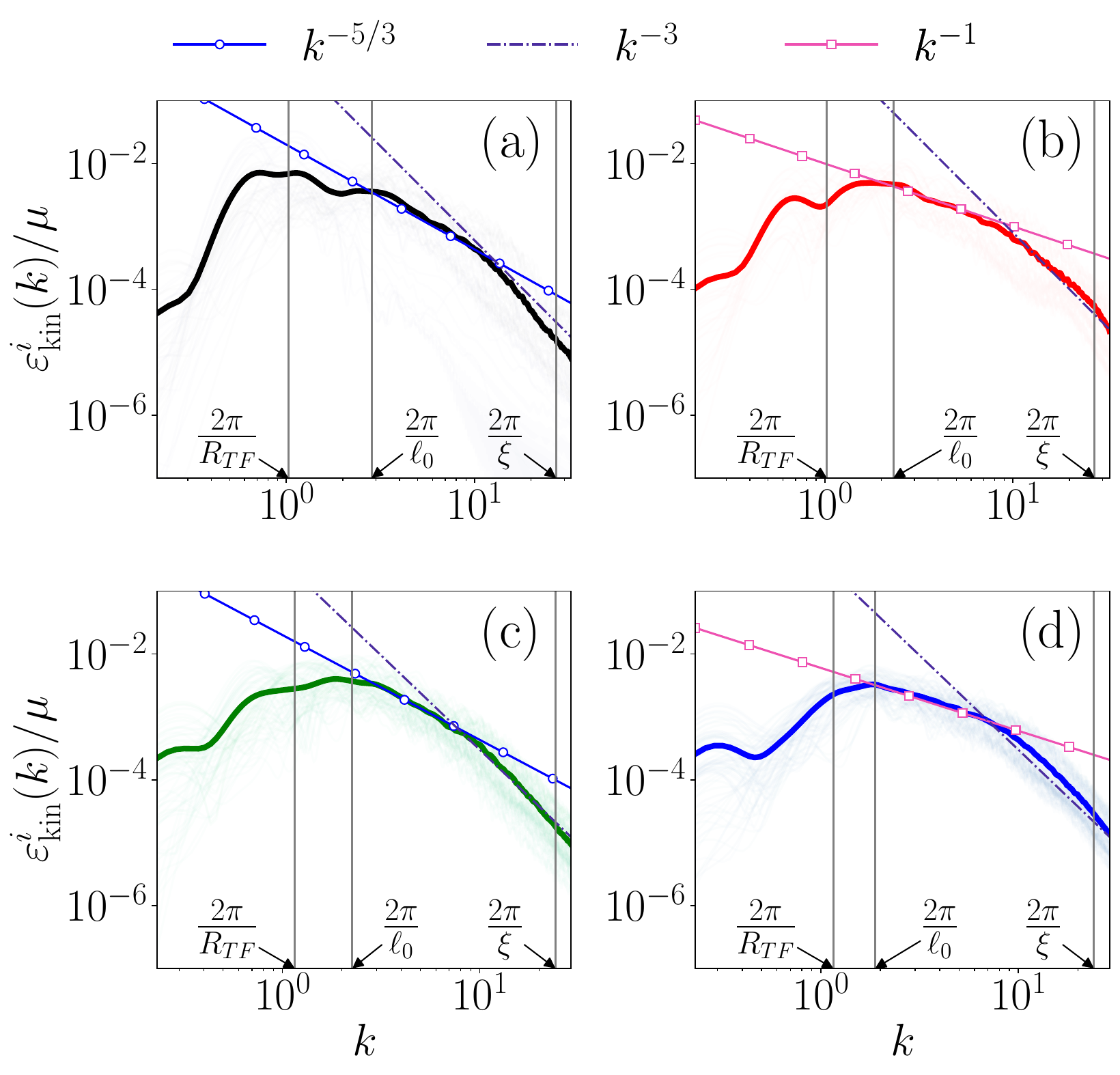}
\caption{Time-averaged incompressible kinetic energy spectra for the non-rotating case, for merging time $t_m=4$ in the time range (a) $t=4$ to $t=14$ (b) $t=14$ to $t=40$ and for merging time $t_m=6$ in the time range (c) $t=6$ to $t=14$ and (d) $t=14$ to $t=40$. Both merging cases initially show $k^{-5/3}$ scaling and later a $k^{-1}$ scaling}.
\label{fig:nr_ikinspectra}
\end{figure}%
We notice that the kinetic energy decays to zero for longer merging times ($t_m=10$, $20$), indicating that the merger is quasi-static for these $t_m$. Further, we find that the total energy gets significantly reduced as merging time increases, indicating a decrease in the excitation energy. This excitation energy depends on the depth of dark solitons, which are shallower for the slower merging cases, as shown in Fig.~\ref{fig:nr_energies}. A point to note is that all these time-varying quantities display oscillations with a frequency of $2.0 \omega$ due to the nature of harmonic confinement. However, these oscillations get reduced when the merging reaches the quasi-static regime where the merger is slow enough not to instigate any dynamical process. These mergers appear to be the same as adiabatic mergers\cite{buchmann2009role}.  Interestingly, the oscillations between potential and kinetic energy suggest that our merger process resembles the turbulence generation mechanism described in the recent work of Middleton-Spencer {\it et al.}\cite{middleton2023}. In their study, an oscillating external potential nucleates vortex and anti-vortex pairs through solitons, which ultimately annihilate to produce density waves. Our system exhibits analogous instabilities following the merging, ultimately culminating in a turbulent state.

Next, to understand the detailed nature of the turbulence for different merging times, in Fig.~\ref{fig:nr_ikinspectra}, we present the incompressible kinetic energy spectra for various merging times. We observe that the merging condensate, except for the $t_m=4$, $6$ cases, does not exhibit any scaling behaviour in the inertial range.  Additionally, the merged condensate immediately after merging ($t_m=4$, $6$) displays a $k^{-5/3}$ scaling for length scales smaller than the inter-vortex distance, $\ell_0$. This differs from the case where turbulence appears in a single condensate, as reported in Ref.~\onlinecite{sivakumar2023energy}. This behaviour can be attributed to the dominance of single vortex dynamics and the lack of sufficient vortices to display a clear Kolmogorov-like cascade of energy. However, at later times (significantly longer than the merging time), the condensate displays $k^{-1}$ scaling at regions smaller than the inter-vortex distance, indicating a Vinen-like cascade\cite{Marino2021, Cidrim2017, sivakumar2023energy}. These scaling behaviours exhibited by the merged condensate are generally short-lived due to the annihilation of vortices in the absence of external mechanisms to generate new vortices, such as rotation. These features are also evident in the absence of a forcing peak at smaller $k$ values, indicating a lack of a proper mechanism to sustain turbulent fluctuations in the merged condensate. To confirm the nature of the turbulence, we performed the merging simulation for a larger grid size. In this case, the incompressible spectra show similar scaling behaviour where both mergers initially display $k^{-5/3}$ and later $k^{-1}$ scalings (see Fig.~\ref{fig:nr_ikinspectra}). Even at larger grid sizes, the presence of $k^{-5/3}$ scaling is not in the range $k < 2\pi/\ell_0$, indicating that a Kolmogorov cascade is not present.

\subsection{Effect of rotation on the dynamics and turbulence state of the merged condensate}
\label{subsec:rotating}

Having a comprehensive insight into the merger of two non-rotating condensates, we now turn our attention to investigating the effect of rotation on the merging dynamics of the condensate, starting with a frequency of ($\Omega_0=0.45$).  %
\begin{figure*}[ht!]
\centering
\includegraphics[width=\linewidth]{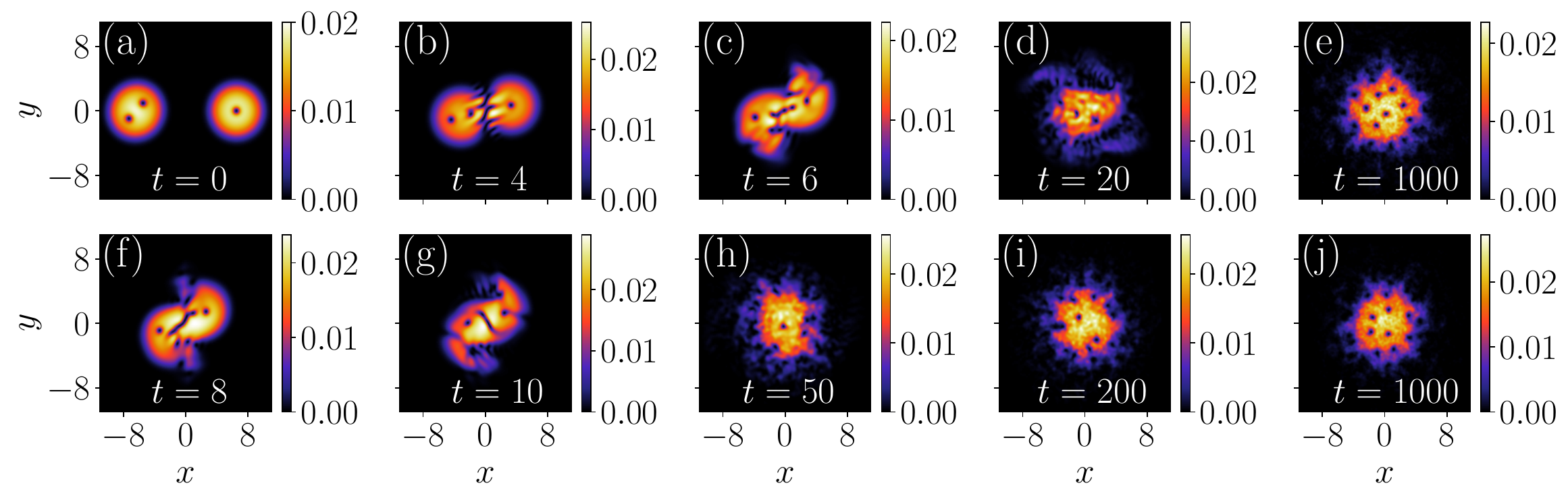}
\caption{Snapshots of the condensate density for the condensates rotating initially at $\Omega_0=0.45$ at different instants of time: (a) $t=0$ and at different instances during the condensate merger for merging times $t_m=6$ [(b)-(e)] and $t_m=10$ [(f)-(j)]. (b) Off-axis collisions forming weaker bent soliton structures, (c) soliton vortex interactions, (d)-(e) formation of single harmonically confined condensate with a vortex lattice, (f)-(g) show off-axis collision with soliton-vortex interaction, and (h)-(j) appearance of the vortex lattice in the merged condensate.}
\label{fig:r45_newdensity}
\end{figure*}
In the rotating merger, we move the traps and their rotating axes towards each other along the $x$-axis while increasing their rotation frequencies at the same rate, resulting in a final merged condensate rotating at $\Omega_f = 2\Omega_0=0.9$.

In Figs. \ref{fig:r45_newdensity}, we present the snapshots of the condensate at different instants of time for $t_m=6 $ [Figs.~\ref{fig:r45_newdensity}(a)-(e)] and $t_m=10$ [Figs.~\ref{fig:r45_newdensity}(f)-(j)]. We notice that the condensates undergo an off-axis merger, deviating from the purely along-the-$x$-axis behaviour seen in the non-rotating case. This off-axis nature, arising from the movement of the rotation axis and trap towards the center of the simulation domain, becomes more pronounced for larger initial separations and shorter merging times ($t_m=6, 10$). In such cases, the condensates can potentially spiral into each other, which strongly influences the nature of the soliton that appears as results of the instability, leading to distortions and orientation at an angle to the $x$-axis due to the presence of rotation. Soliton-vortex interactions  also become more probable for slower mergers (see Fig.~\ref{fig:r45_newdensity} bottom panel), but compared to the situation as discussed in Ref.~\onlinecite{Kanai2019}, these interactions do not affect the vortex numbers or angular momenta in presence of the rotation. %
\begin{figure}[!ht]
\centering
\includegraphics[width=\linewidth]{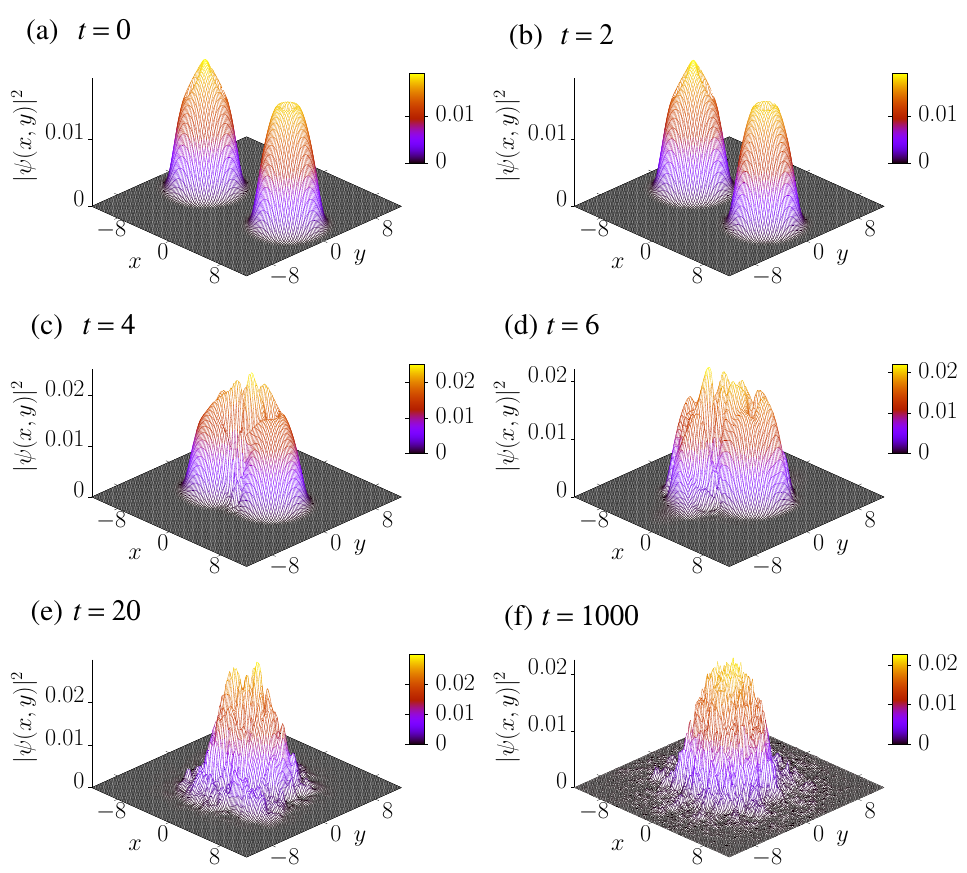}
\caption{
Three-dimensional visualizations of the condensate density surface at (a) $t=0$ [cf. Fig.~\ref{fig:r45_newdensity}(a)], (b) $t=2$, and various time instances, (c) $t=4$, (d) $t = 6$, (e) $t = 20$ and (f) $t=1000$, corresponding to the merging time $t_m=6$ [cf. Fig.~\ref{fig:r45_newdensity}(b) - \ref{fig:r45_newdensity}(e)].}
\label{fig:r45:surf}
\end{figure}
Figure~\ref{fig:r45:surf} provides a comprehensive visualization of the surface plots of condensate densities for rotational merger with $t_m=6$ corresponding to the time points depicted in Figs.~\ref{fig:r45_newdensity}(a)-\ref{fig:r45_newdensity}(e), including an intermediate time $t=2$ in Fig.~\ref{fig:r45:surf}(b).

\begin{figure}[!htp]
\centering
\includegraphics[width=\linewidth]{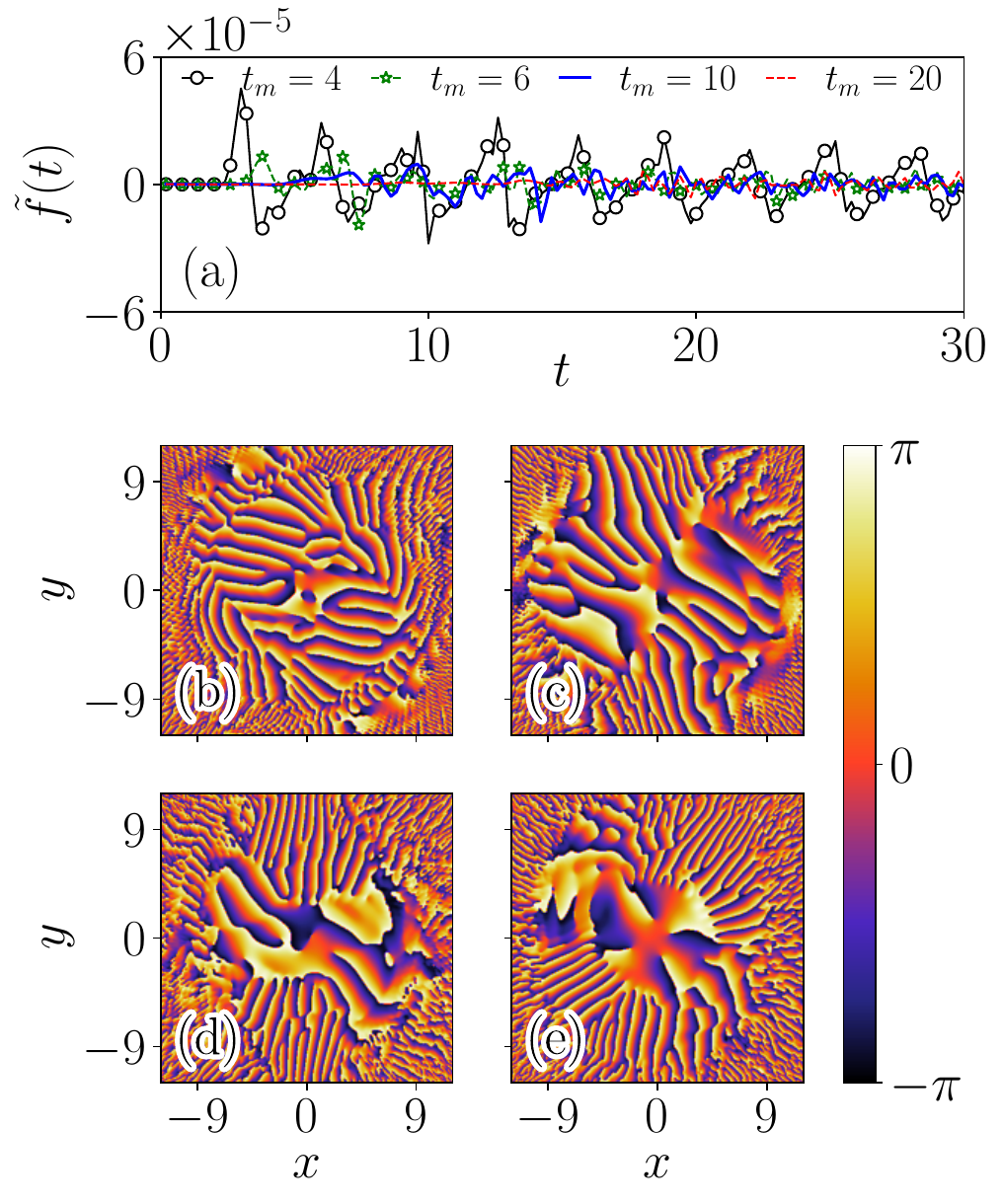}
\caption{(a) Temporal evolution of the forcing term $f$ with rotating case for different merging times: $t_m=4$ (black solid line with empty squares), $t_m=6$ (green dashed line with circles), $t_m=10$ (blue solid line), and $t_m=20$ (red dashed line). Condensate phase profiles at the time of merger for various merging times: (b) $t_m=4$, (c) $t_m=6$, (d) $t_m=10$ and (e) $t_m=20$.}
\label{fig:r45_forcing}
\end{figure}
Next, we show the temporal evolution of the forcing of the merged condensate in Fig.~\ref{fig:r45_forcing}(a) for different $t_m$, which follows a similar pattern to that of the non-rotating case but has a significantly smaller forcing amplitude due to the presence of shallower solitons that exhibit significant bending. In this case, the forcing term lies within the interval $\sim [-10^{-5}, 10^{-5}]$ with the standard deviation of approximately $10^{-8}$ for $t_m=4$ and $t_m=6$, and that of the order of $10^{-9}$ for $t_m=10$ and $t_m=20$. Notably, the amplitude decays much slower when compared to the non-rotating case and persists for longer times due to the presence of rotational effects. The long correlation for the forcing can be further confirmed through the slow decay of the compressible energy peak as illustrated in Fig.~\ref{fig:r45_compden} (see Appendix \ref{appendix_a}). The phase profiles in Figs.~\ref{fig:r45_forcing}(b)-(e) illustrate the nature of the soliton at the instant when merging takes place for $t_m=4$, $6$, $10$, and $20$ respectively for the rotating condensate. For small merging times ($t_m=4$, $6$), we notice the presence of a large number of solitons, which are quite evident from the discontinuous phase jumps [see  Fig.~\ref{fig:r45_forcing}(b),(c)]. Conversely, for  $t_m=10$, $20$, the phase profile is smoother, indicating the presence of fewer solitons and the emergence of vortices [see Figs.~\ref{fig:r45_forcing}(d)-(e)]. Similar to the non-rotating case, we can relate the large velocity jump to the high velocity of the soliton, while the smooth phase corresponds to the soliton moving with the sound velocity\cite{Kanai2019}. %
\begin{figure}[!htp]
\centering
\includegraphics[width=\linewidth]{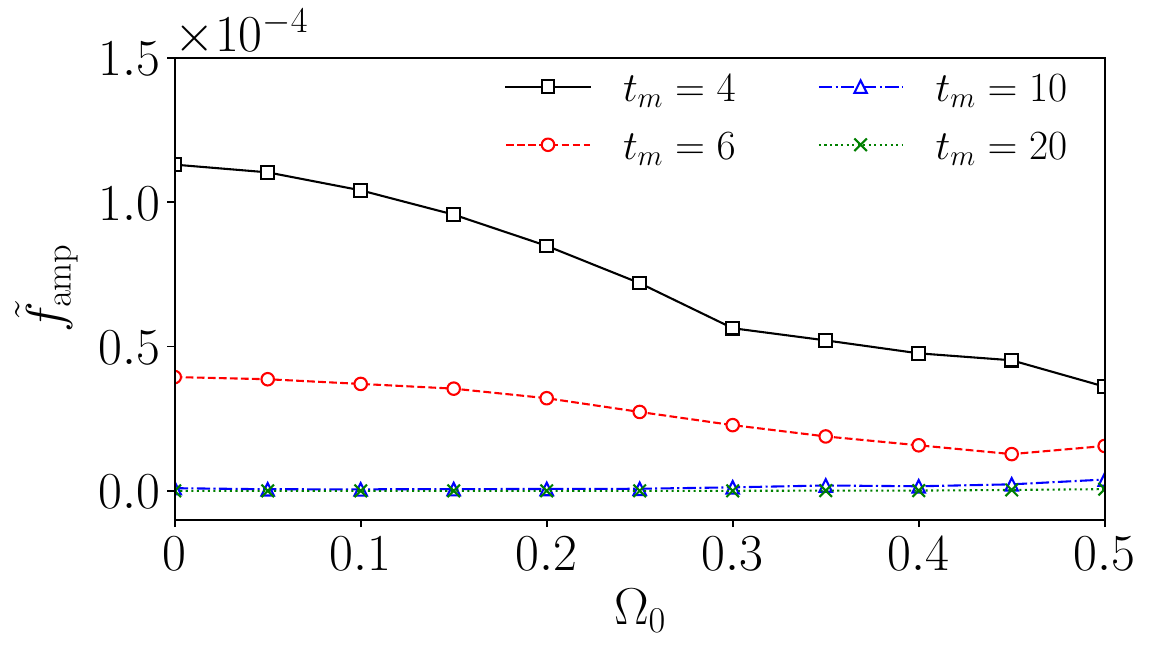}
\caption{Variation of forcing amplitude against the rotating frequency for different merging times of $t_m=4$, $t_m=6$, $t_m=10$, and $t_m=20$.}
\label{fig:forcingamp}
\end{figure}
To characterize the effect of the rotation frequency on the force amplitude, which in turn affects the structure and dynamics of the soliton. In Fig.~\ref{fig:forcingamp}, we depict the variation of the force amplitude ($\tilde{f}_{amp}$) with the angular frequency  ($\Omega_0$) of the rotation for different $t_m$. We find that for small merging times ($t_m=4$, $6$), the force amplitude decreases with an increase in $\Omega_0$ .However, for longer merging times ($t_m=10$, $20$), the force amplitude is close to zero and independent of $\Omega_0$. 

\begin{figure}[!ht]
\centering
\includegraphics[width=\linewidth]{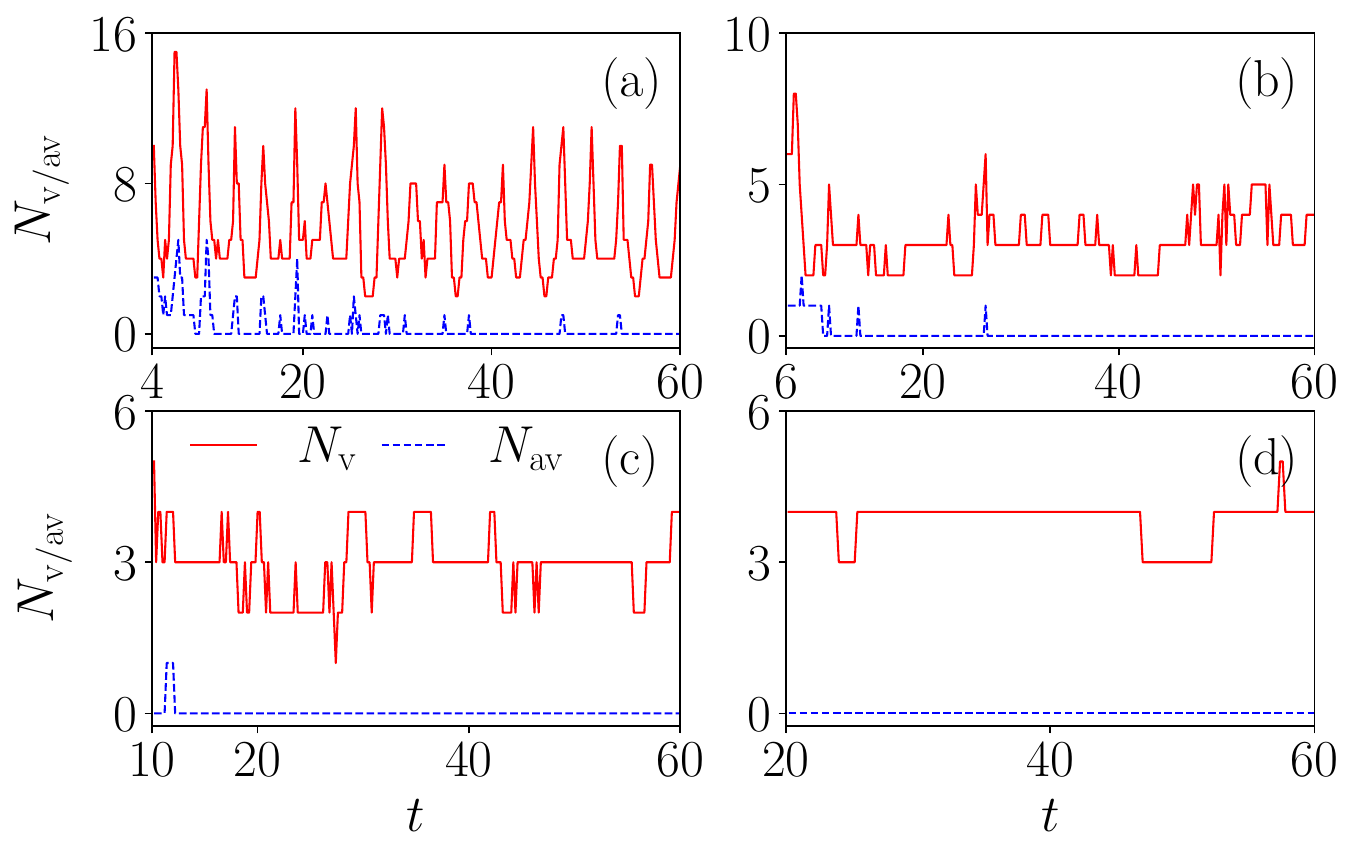}
\caption{Evolution of vortex ($N_{\mathrm{v}}$) and anti-vortex ($N_{\mathrm{av}}$) numbers within the Thomas-Fermi radius of the condensate ($R_{\mathrm{TF}}$)  for (a) $t_m=4$ (b) $t_m=6$ (c) $t_m=10$ (d) $t_m=20$ for rotating case with $\Omega_0=0.45$.}
\label{fig:r45_vortex_number}
\end{figure}
There is a significantly higher vortex count at shorter merging times. This phenomenon arises from the formation and subsequent decay of solitons into vortex-antivortex pairs, contributing to the vortex-population in cases where $t_m=4,6$. For longer merging times, the likelihood of soliton formation diminishes, resulting in a decrease in the contribution of soliton decay to the vortex population. Additionally, we find that the rotational effect causes the vortex population to surpass the anti-vortex population which is quite evident from the evolution of vortex numbers as shown in Figure \ref{fig:r45_vortex_number}. For faster mergers in Fig.~\ref{fig:r45_vortex_number}(a), the vortex number oscillates at $2 \omega$ due to density oscillations. However, we witness a significant reduction of this behaviour for slower mergers in Figs. \ref{fig:r45_vortex_number}(c)-(d), following a trend similar to the previous case.

\begin{figure}[!ht]
\centering
\includegraphics[width=\linewidth]{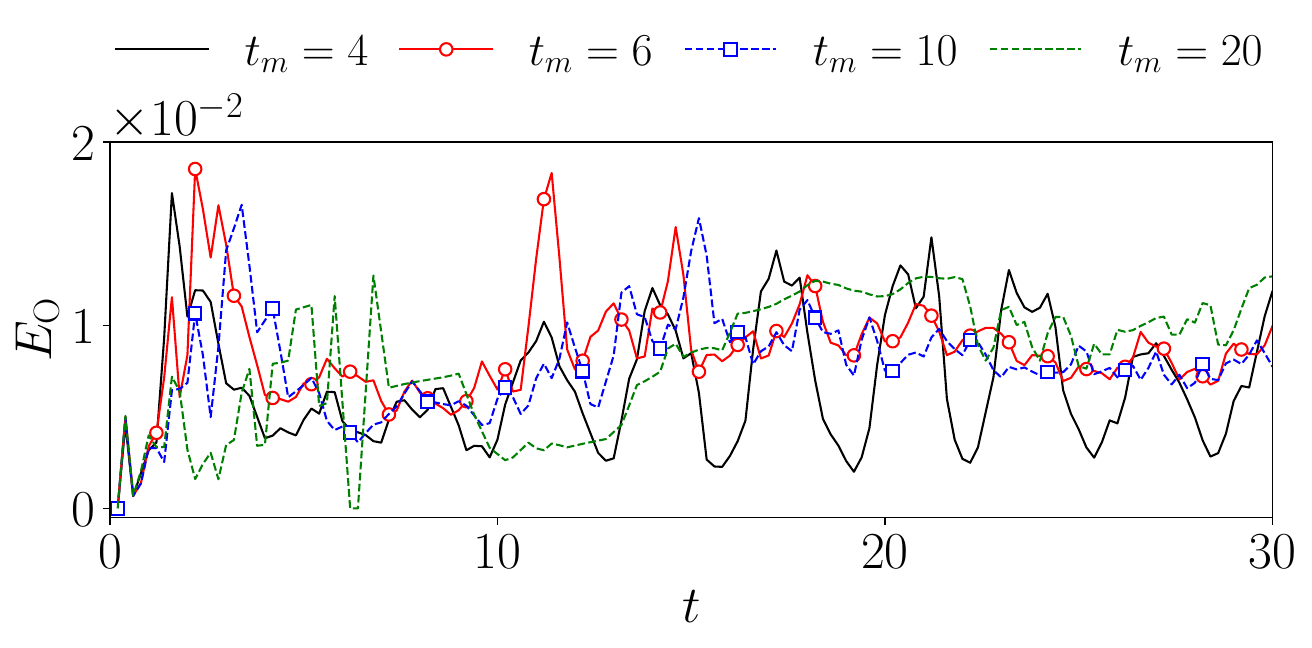}
\caption{Evolution of Onsager energy $E_{\mathrm O}$ with respect to time for the non-rotating case for merging times $t_m=4$ (black solid line), $t_m=6$ (red solid line with empty circles), $t_m=10$ (blue dashed line), $t_m=20$ (green dashed line with empty squares).}
\label{fig:r45_onsager}
\end{figure}
The incompressible kinetic energy evolution exhibits a decreasing trend concerning merging time. In contrast to the non-rotating scenario, the incompressible energy does not quickly settle down. Instead, it persists indicating a continuous production of vortices. Examining the compressible kinetic energy reveals a trend similar to that of the non-rotating case. However, for longer merging times ($t_m=20$), there is a persistent and constant compressible energy, while in the non-rotating case, the compressible energy approaches zero at these times. This discrepancy is primarily due to the presence of rotational forcing, which not only generates more vortices but also enhances the production of compressible sound waves. The Onsager energy profile for different merging times in the rotating case is quite different, as shown in Fig.~\ref{fig:r45_onsager}. In the rotating case, vortices are not only formed via decaying solitons but also due to the presence of rotation. From Fig.~\ref{fig:r45_onsager} one can see that vortices formed via rotation have more Onsager energy when compared to the ones formed via solitons. The lowest Onsager energy for $t_m=4$ (black line) can be attributed to the significant increase in the number of vortices formed due to snake instability, which lowers the overall Onsager energy in the condensate despite the presence of vortices formed due to rotation.

\begin{figure}[!htp]
\centering
\includegraphics[width=\linewidth]{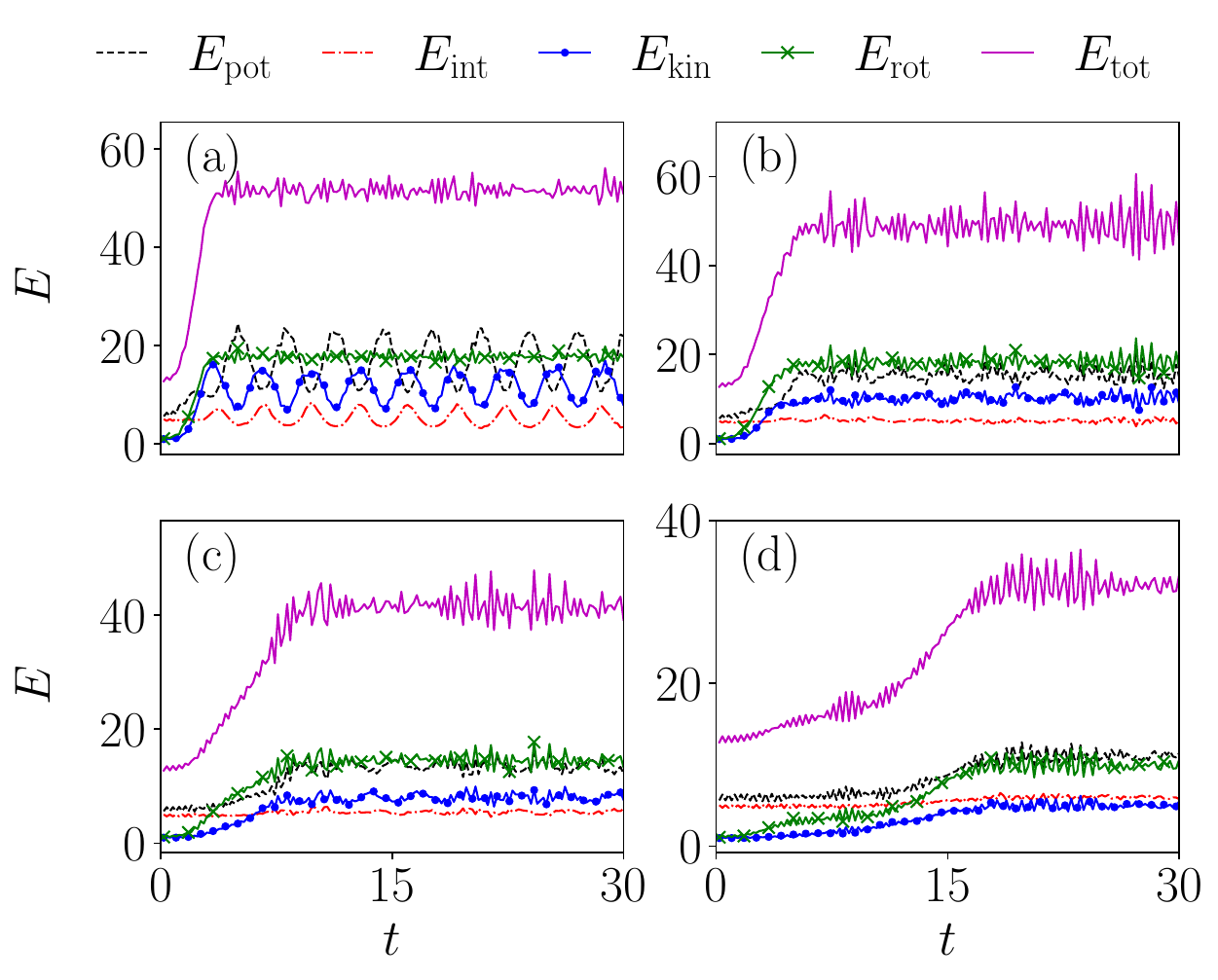}
\caption{Evolution of different components of condensate energies with respect to time for (a) $t_m=4$, (b) $t_m=6$, (c) $t_m=10$, and (d) $t_m=20$ for rotating case with $\Omega_0=0.45$.}
\label{fig:r45_energies}
\end{figure}

In Fig.~\ref{fig:r45_energies}(a)-(d), we show the various energy components, $E_{pot}$, $E_{int}$, $E_{kin}$, $E_{rot}$ for different merging times, i.e., $t_m=4$, $6$, $10$, and $20$, respectively, keeping the rotation frequency fixed at $\Omega_0=0.45$. %
Similar to the non-rotating case, all energy components (except for rotation energy) exhibit oscillations consistent with the trap frequency for shorter merging times. However, unlike the non-rotating case, the amplitude of these oscillations damps out quickly [see Figs.~\ref{fig:r45_energies}(b)-(d)]. The kinetic energy decreases with increasing merging time, but unlike the non-rotating merger, it does not go to zero for the $t_m=10$ and $t_m=20$ cases.

\begin{figure}[!ht]
\centering
\includegraphics[width=\linewidth]{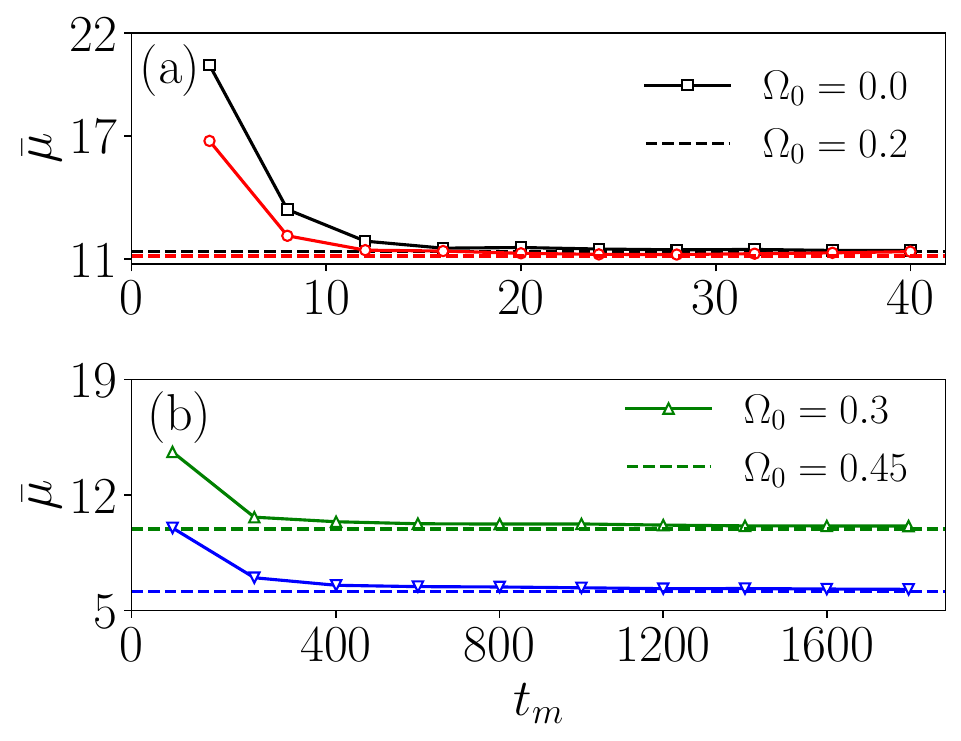}
\caption{Evolution of chemical potential (averaged over $3000$ time units from the time of merger) of the merged condensate with respect to different merging times for (a) $\Omega_0=0.0$ (black line with empty squares) and $\Omega_0=0.2$, (red line with empty circles) (b) $\Omega_0=0.3$ (green line with empty triangles) and $\Omega_0=0.45$ (blue line with empty triangles) frequencies, respectively. The dashed horizontal lines represent the ground state chemical potential for the merged condensate rotating at $2\Omega_0$.}
\label{fig:ground_states}
\end{figure}

\begin{figure}[ht!]
\centering
\includegraphics[width=\linewidth]{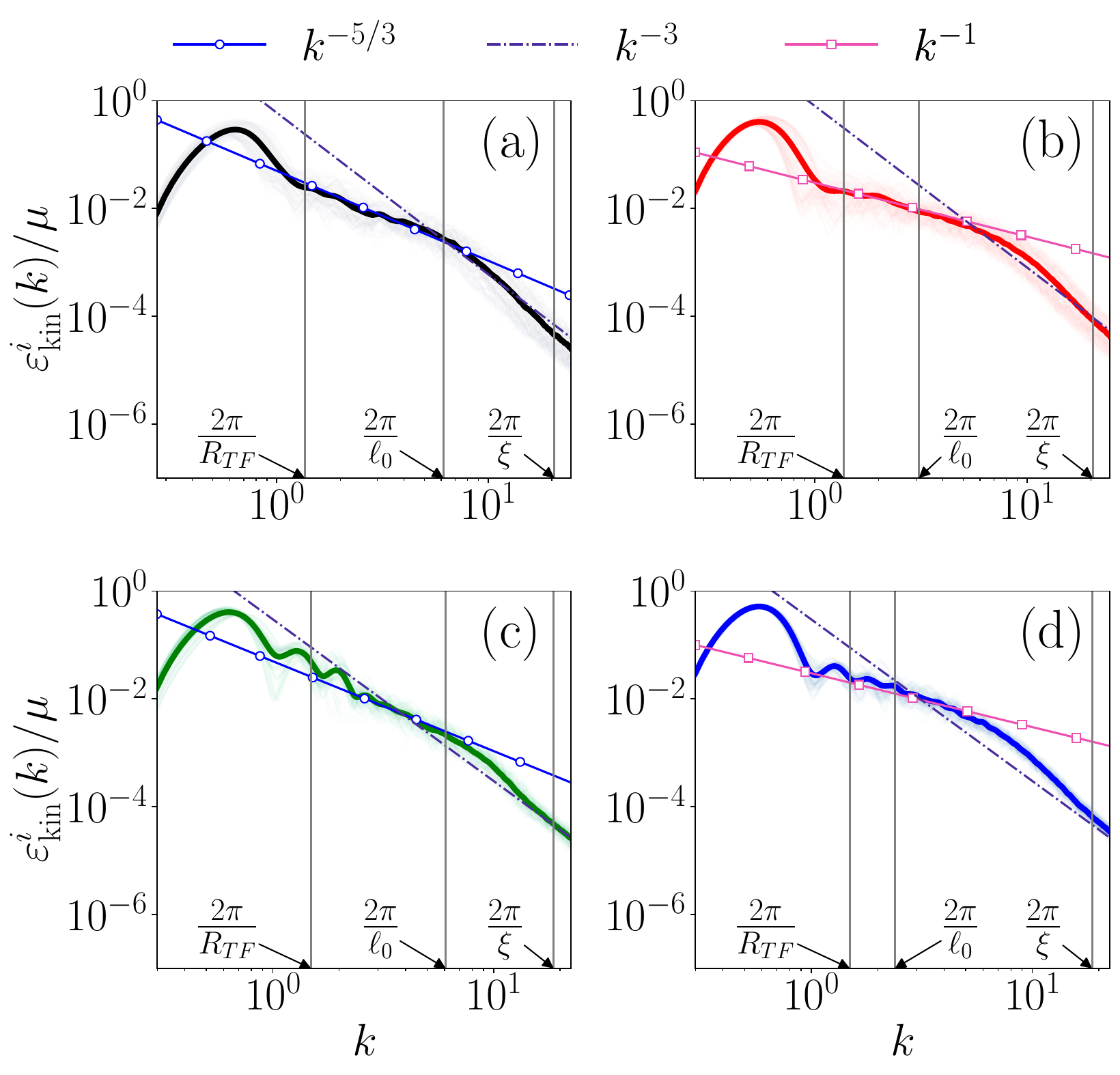}
\caption{Time-averaged incompressible kinetic energy spectra for the rotating case ($\Omega_0=0.45$) for different merging time (a,b) for $t_m=4$ and (c,d) for $t_m=6$.  (a) represents the averaged spectra in the range of time interval $t=4$ to $t=24$ and (b) $t=24$ to $t=44$. For merging time $t_m=6$ in the time range (c) $t=6$ to $t=26$ and (d) $t=26$ to $t=46$. Both cases initially showing $k^{-5/3}$ and later a $k^{-1}$ scaling.} 
\label{fig:r45_ikinspectra}
\end{figure}

Achieving an adiabatic merger in the condensate depends on both the merging time ($t_m$) and the rotation frequency ($\Omega_0$), as shown in Fig.~\ref{fig:ground_states}. Condensates rotating at frequency $\Omega_0$ merge to form a condensate rotating at $2\Omega_0$. Sufficiently long merging times ensure that this merged condensate reaches its ground state. Notably, for both non-rotating condensates and those at low rotation frequencies [black line with empty squares in Fig.~\ref{fig:ground_states}(a)], the ground state is achieved with merging times of $t_m \sim 20$. In Figs.~\ref{fig:ground_states}(b), as the rotation frequency increases, longer merging times are required for the final condensate to approach its ground state. Even without soliton decay, trap movement can create compressible excitations. These excitations persist longer at higher rotation frequencies, as rotational forcing prevents them from dissipating, thus requiring much slower mergers to suppress them.

The incompressible kinetic energy spectra for the rotating case exhibit similar scaling behaviour to that of the non-rotating case (see Fig.~\ref{fig:r45_ikinspectra}). It is important to note that a forcing peak emerges due to rotation at $k < 2\pi/R$ gradually diminishes for longer merging times. For cases with $t_m=4$ and $t_m=6$, a short-lived $k^{-5/3}$ scaling is initially observed, which later transforms into a $k^{-1}$ scaling, corresponding to the formation of vortex-rings (see Ref. \onlinecite{barenghi2023types}). These vortex rings are formed in the soliton decay process through snake instability, as illustrated in Ref. \onlinecite{verma2017generation}.

In contrast to the nonrotating case, where no scaling is observed for $t_m=10$ and $t_m=20$, the rotating case exhibits $k^{-5/3}$ scaling, albeit for a relatively shorter $k$ range initially. However, the $k^{-1}$ scaling persists for the $t_m=10$ case but not for the $t_m=20$ case, as no soliton is formed to decay into vortex-rings. In the non-rotating scenario, the scaling behaviour of the incompressible spectra disappears due to vortex annihilation. Also, in contrast to the non-rotating case, the $k^{-5/3}$ scaling appears consistently for length scales larger than inter-vortex distance $\ell_0$, where collective vortex dynamics dominate and hence display Kolmogorov cascade. This is due to rotational effects that introduce a significant number of vortices into the condensate, enabling their collective behaviour. This is possible for frequencies above critical frequency $\Omega_0 > \Omega_{\mathrm{c}}$\cite{kumar2019c}, which in our case is $\Omega_{\mathrm{c}}\sim 0.3$. Similarly, the $k^{-1}$ scaling also appears at smaller length scales where single-vortex dynamics dominate, clearly indicating Vinen-like turbulence. In the presence of sustained rotation, we observe a more pronounced Kolmogorov cascade of $k^{-5/3}$ at smaller $k$ regions after longer times. This phenomenon closely resembles the later stage evolution of rotating condensates, as previously studied in Ref. \onlinecite{sivakumar2023energy}.

\section{Conclusion}
\label{sec:conclusion}

We have numerically investigated the merging dynamics of Bose-Einstein condensates with random phases confined in harmonic traps, analyzing the impact of rotation on the process. Initially, we studied mergers without rotation for various merging times. All exhibited consistent scaling behaviour in the incompressible kinetic energy spectra. The density profiles at different merging times showed the formation of dark solitons due to interference patterns generated by condensate overlap. In particular, interference fringes closer to the minima of the final trap persisted longer than those near the boundaries. These dark solitons exhibited bending and subsequently decayed into vortex pairs via snake instabilities. Vortex-antivortex pairs eventually annihilated after multiple rebounds within the trap. However, while observing these features for all merging times ($t_m = 4$, $6$, $10$, and $20$), the nature of the dark solitons, their instability-driven behaviour, and the phenomenon's timeframe varied significantly. Increasing the merging time resulted in a decrease in the depth and size of the central dark soliton.  Since soliton depth influences phase excitations, slower merging led to the attainment of the BEC's ground state. This depth decrease was confirmed by analyzing the time-dependent variation of the forcing term. Furthermore, the vortex number and incompressible kinetic energy density profiles provided temporal evidence of vortex formation and annihilation. The compressible kinetic energy, indicating sound waves from phase excitations, also showed a decreasing trend and almost disappeared for longer merging times.  In the beginning, the incompressible kinetic energy spectra exhibited $k^{-5/3}$ but only at length scales where single vortex dynamics dominate, hence ruling out the Kolmogorov cascade. Later, for a short period, they transitioned to a $k^{-1}$ scaling attributed to vortex rings formed from soliton breaking via the snake instability\cite{barenghi2023types}. However, the absence of forcing in the nonrotating case prevented these scaling laws from persisting beyond vortex annihilation. The formation of vortices through merging BECs with random initial phases, even in the absence of rotation, agrees with conclusions drawn from Refs.~\onlinecite{buchmann2009role, scherer2007vortex}. In the case of both rotating and non-rotating mergers, the interaction, potential, and kinetic energy components execute oscillations consistent with frequency $2.0\omega$. Such oscillations (generated by oscillating the traps externally) are shown to create turbulence via solitons and vortex pair annihilations\cite{middleton2023}, similar to our results presented here where these oscillations are triggered by fast collisions. Conversely, for slower mergers, these oscillations get subsided, thus reducing the turbulent effects. These effects are corroborated by the scaling laws observed in the energy spectra for shorter merging times and the absence thereof for longer merging times.

With rotation, the merging condensates display contrasting behaviour for identical merging times, consistent with observations without rotation. The evolution of the density indicates that the condensates merge off the $x$-axis despite the model restricting translational motion strictly along the $x$-axis, which results in them experiencing each other's rotational effects as they approach. It significantly weakens the resultant dark solitons and orients with an angle to the $x$-axis. These solitons also appear to interact with the preexisting and newly formed vortices. After merging, their rotating frequencies combine, resulting in the final condensate rotating at twice the initial frequency of the individual condensates The significantly lower amplitude of the time-dependent forcing term confirms the weak nature of the solitons compared to the non-rotating case. The incompressible kinetic energy profiles indicate continuous vortex production due to sustained rotation and compressible turbulence generation. The number of vortices exceeds the anti-vortex number due to the rotation's impact on the vortex population.

Although compressible kinetic energy exhibits a decreasing trend, it does not necessarily approach zero as rapidly for longer merging times, as observed in the non-rotating case. This finding, in conjunction with the average chemical potential plot, indicates that an adiabatic merger is feasible only for merging times exceeding $t_m=1000$, unlike non-rotating cases where the merged condensate reaches the ground state by $t_m=20$. While the resulting soliton is shallower and has a weaker influence on phase excitations, the rotational forcing significantly impacts the sound waves produced, which persist without dissipation. Notably, during adiabatic mergers in the rotating case, the quasi-static nature of the merger substantially reduces off-axis merging.

The incompressible kinetic energy spectra exhibit a scaling behaviour similar to that of the non-rotating case. However, these spectra display a forcing peak at $k < 2\pi/R$. Initially, the scaling range extends to length scales longer than the intervortex length $k < 2\pi/\ell_0$ indicating the onset of the Kolmogorov cascade and at later times displays Vinen-like turbulence. With sustained rotation over a longer duration, we observe spectra similar to those presented in our previous work for the $\Omega_f=1$ case\cite{sivakumar2023energy}, where the Kolmogorov cascade is observed at smaller $k$ values. The presence of rotation significantly enhances vortex production, affecting turbulence in the merged condensate. For rotation frequencies below $0.3$, although the incompressible spectra show a $k^{-5/3}$ scaling, this scaling range exists only for wavenumbers larger than the intervortex distance ($2\pi/\ell_0$). This indicates the absence of a Kolmogorov energy cascade due to the insufficient number of vortices. Only with stronger rotation (frequencies above $0.3$) and increased vortex production do we observe scaling for wavenumbers below $2\pi/\ell_0$, that exhibit Kolmogorov-like cascade.

On the behaviour of the energy oscillations for the rotating case, it is quite evident that they are not as strong as those observed with the non-rotating case and decay relatively quickly for longer merging times. So, the distribution of the compressible energy is not granulated like in the non-rotating case (see. Appendix figure \ref{fig:nr_compden} but appear more clumped in the form of strands (see. Appendix figure \ref{fig:r45_compden}). Despite weakened compressible turbulence, the incompressible kinetic energy spectra show a stronger energy cascade relative to its non-rotating counterpart, which one could attribute to the vortex production augmented due to the presence of rotation.

Barenghi {\it et al.}\cite{barenghi2023types} have highlighted the complex dynamics of quantum turbulence into three primary types: Kolmogorov turbulence, Vinen quantum turbulence, and strong quantum turbulence based on their underlying mechanisms and comparisons with classical turbulence. In our work, we observed that rapid mergers can induce turbulent instability, causing the condensate to oscillate and nucleate solitons that migrate to the periphery, reflecting and colliding to form vortex-antivortex pairs.  These collisions produce sound waves and, in conjunction with density fluctuations, contribute to intricate, turbulent fluid dynamics. Although rotational forcing generates fewer sound waves, it stimulates vortex production and breakdown, resulting in a Kolmogorov cascade characteristic of strong quantum turbulence\cite{middleton2023, barenghi2023types}. As vortex pairs annihilate and solitons decay, the system transitions to a Vinen turbulence regime.

We have also carried out the simulations at various rotating frequencies and identified trends. At lower frequencies, both the scaling duration and ranges are significantly shorter, while the depth of the dark soliton is larger compared to higher rotation frequencies. Additionally, off-axis merging is less pronounced at lower frequencies. Despite these trends, the rotational case does not exhibit fundamental differences across various rotating frequencies to the same extent as the non-rotating case.

Finally, it is worthwhile to discuss the experimental feasibility of verifying the instability and turbulence observed in this study. Given the trap confinement parameters described in Sec. \ref{sec:model}, condensates of approximately $1 \mu\mbox{m}$ size are expected to merge within milliseconds, potentially leading to turbulent behaviour. While rotational mergers are challenging to tackle due to the spiralling nature of the condensates, experimental verification of turbulence has been achieved for non-rotating superfluids\cite{Escartin2019}. One possible approach to triggering mergers is by gradually reducing the central barrier in a double-well potential, causing the condensates to combine into a single well. However, this method is not suitable for rotational mergers because the condensates collide off the merging axis. To overcome this issue, an additional velocity could be imparted to the condensates along their individual rotational motion around the $z$-axis.

\acknowledgments
The works of A.S. and P.M. are supported by MoE RUSA 2.0 (Bharathidasan University - Physical Sciences).
\counterwithin{figure}{section}

\appendix

\section{Compressible kinetic energies}

\label{appendix_a}

\begin{figure}[!ht]%
    \centering
    \includegraphics[width=\linewidth]{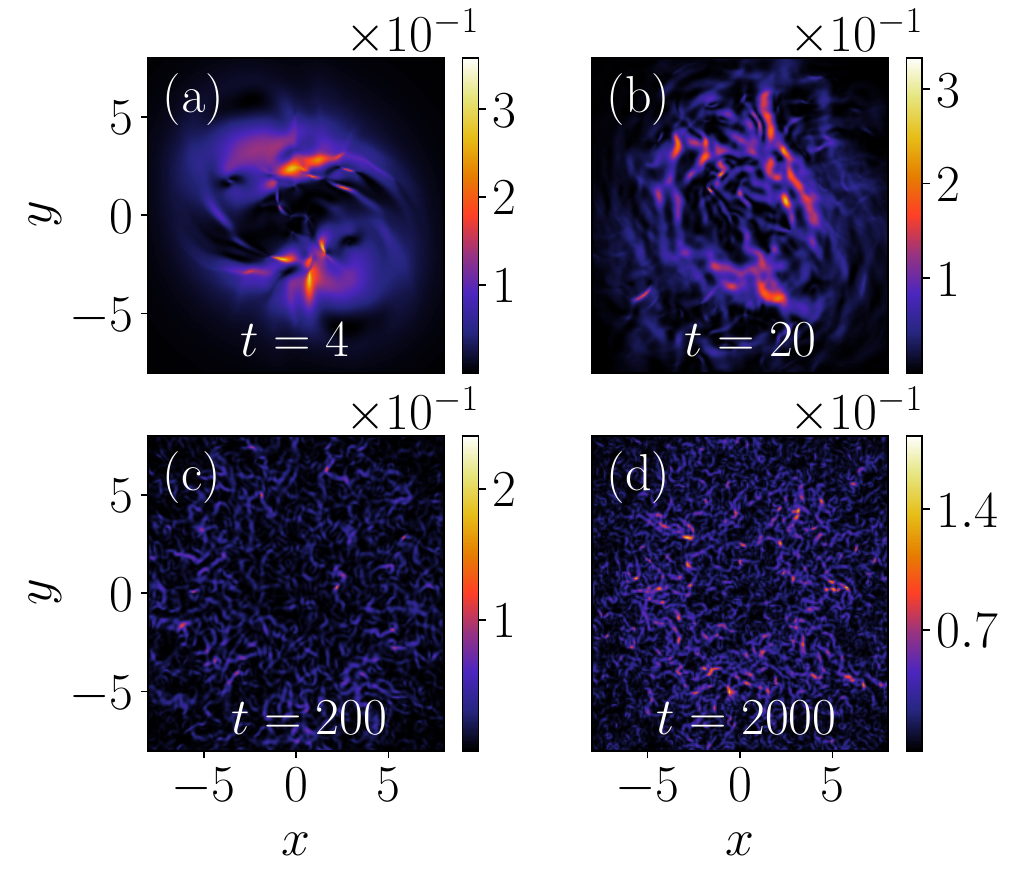}
    \caption{Plots showing the time evolution of compressible energy density for rotating condensates with merging time $t_m=4$ at different instants of time: (a) $t = 4$, (b) $t = 20$, (c) $t = 200$, and (d) $t = 2000$.}
    \label{fig:r45_compden}
\end{figure}

In this appendix, we present the spatial configuration of the compressible kinetic energy for $t_m=4$, which is useful to characterize the different sorts of instabilities appearing after the merging leading to the turbulent state for both rotating as well as non-rotating condensates. %
\begin{figure}[!ht]
    \centering
    \includegraphics[width=\linewidth]{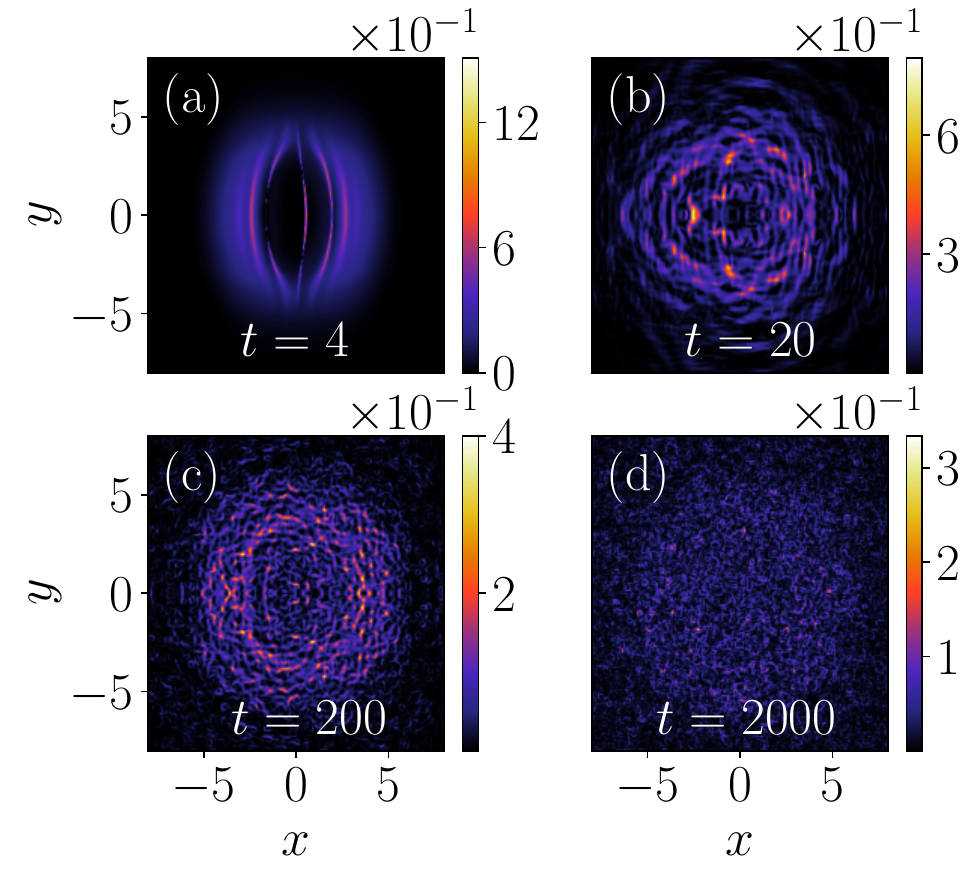}
    \caption{Plots illustrating the time evolution of compressible energy density for non-rotating condensates with merging time $t_m=4$ at different instants of time: (a) $t = 4$, (b) $t = 20$, (c) $t = 200$, and (d) $t = 2000$.}
    \label{fig:nr_compden}
\end{figure}%
Upon observing the time evolution of the forcing term, we notice that the forcing amplitude decays relatively quickly for the non-rotating case compared to the rotating one. Additionally, the forcing experienced in the condensate manifests as compressible energy generated in the merged condensate. We highlight this by analyzing the spatial density profiles of compressible kinetic energy. For the non-rotating case, we observe that the dark solitons contribute to the compressible energy density [see Fig.~\ref{fig:nr_compden}(a)]. Eventually, the energy becomes more evenly distributed and finally granulates at longer times [see Fig.~\ref{fig:nr_compden}(b)-(d)]. The decreasing amplitude of the energy density and the initial accumulation of compressible energy in solitons correlates with the evolution of the forcing term. The energy density for the rotating case further confirms this correlation, as the energy amplitude is significantly lower and distributed over a wider soliton. Similar to the forcing amplitude decreasing at a slower rate, the compressible energy peaks also do not decay quite as fast as in the non-rotating case within the same time interval (see Fig.~\ref{fig:r45_compden}). %
It is also noteworthy that, compared to the non-rotating case, the compressible energy is distributed as strands and is more clumped at longer times. This spatial profile resembles the long-term evolution of compressible kinetic energy shown in rotational quantum turbulence\cite{sivakumar2023energy}.


%

\end{document}